\documentclass[isre,nonblindrev]{informs3} 

\DoubleSpacedXI

\usepackage{natbib}
 \bibpunct[, ]{(}{)}{,}{a}{}{,}%
 \def\bibfont{\small}%

\usepackage{float} 
\usepackage{booktabs} 
\usepackage{bbm} 
\usepackage{bm}
\usepackage{longtable}
\usepackage{booktabs}
\usepackage{lscape} 
\usepackage{afterpage}
\usepackage{multibib} 
\newcites{app}{Appendix References}

\TheoremsNumberedThrough     
\ECRepeatTheorems

\EquationsNumberedThrough    

\MANUSCRIPTNO{ISRE-0000-0000.00}

\renewcommand{\theARTICLETOP}{}
\begin{document}


\RUNAUTHOR{Ma et al.} 

\RUNTITLE{Learning to Adopt Generative AI}

\TITLE{Learning to Adopt Generative AI}

\ARTICLEAUTHORS{%
\AUTHOR{Lijia Ma}
\AFF{Belk College of Business, University of North Carolina at Charlotte, Charlotte, NC 28223, \EMAIL{lijiam@uw.edu}} 
\AUTHOR{Xingchen (Cedric) Xu}
\AFF{Michael G. Foster School of Business, University of Washington, Seattle, WA 98195, \EMAIL{xcxu21@uw.edu}} 
\AUTHOR{Yumei He}
\AFF{Freeman School of Business, Tulane University, New Orleans, LA 70118, \EMAIL{yhe17@tulane.edu}} 
\AUTHOR{Yong Tan}
\AFF{Michael G. Foster School of Business, University of Washington, Seattle, WA 98195, \EMAIL{ytan@uw.edu}}
} 

\ABSTRACT{%
Recent advancements in generative AI, such as ChatGPT, have dramatically transformed how people access information. Despite its powerful capabilities, the benefits it provides may not be equally distributed among individuals, a phenomenon referred to as the \textit{digital divide}. Building upon prior literature, we propose two forms of \textit{digital divide} in the generative AI adoption process: (i) the \textit{learning divide}, capturing individuals' heterogeneous abilities to update their perceived utility of ChatGPT; and (ii) the \textit{utility divide}, representing differences in individuals' actual utility derived from per use of ChatGPT. To evaluate these two divides, we develop a Bayesian learning model that incorporates heterogeneities in both the utility and signal functions. Leveraging a large-scale clickstream dataset, we estimate the model and find significant \textit{learning and utility divides} across various social characteristics. Interestingly, individuals without any college education, non-white individuals, and those with lower English literacy derive larger utility gains from ChatGPT, yet update their beliefs about its utility at a slower rate. Furthermore, males, younger individuals, and those in occupations with greater exposure to generative AI not only obtain higher utility per use from ChatGPT but also learn about its utility more rapidly. Besides, we document a phenomenon termed the \textit{belief trap}, wherein users underestimate ChatGPT's utility, opt not to use the tool, and thereby lack new experiences to update their perceptions, leading to continued underutilization. Our simulation further demonstrates that the \textit{learning divide} can significantly affect the probability of falling into the \textit{belief trap}, another form of the digital divide in adoption outcomes (i.e., \textit{outcome divide}); however, offering training programs can alleviate the \textit{belief trap} and mitigate the divide. Our findings contribute to the literature on the digital divide, AI adoption, and Bayesian learning.

}%


\KEYWORDS{Generative AI, AI Adoption, Digital Divide, Bayesian Learning, Human-AI Interaction}

\maketitle

%

\newpage
\section{Introduction}

\subsection{Motivation}

Search engines have long been the primary means by which individuals access information; over 93\% of internet activities start with a search query\footnote{See \url{https://www.webfx.com/seo/statistics/}}. For instance, Google, which dominates the market with approximately 90\% share, processes about 2 trillion searches each year\footnote{See \url{https://blog.hubspot.com/marketing/google-search-statistics}}. However, the recent emergence of large language models (LLMs) offers a new means to access these resources. Trained on extensive datasets \citep{zhang2024regurgitative}, LLM-backboned generative AI, such as ChatGPT, can provide natural language responses to a wide range of user queries \citep{wang2023human}. The general-purpose nature of generative AI powered by LLMs \citep{morris2023levels}, combined with their vast knowledge base, positions them as a compelling alternative to search engines \citep{xu2023chatgpt}.

Given the significant potential of generative AI, one might anticipate widespread benefits. However, these potential advantages may still be constrained by the ``digital divide," which refers to the disparities in access to, capability with, and effective use of information and communication technologies (ICT) among individuals from diverse backgrounds \citep{vassilakopoulou2023bridging, oecd2001understanding}. Understanding and addressing the digital divide is crucial when aiming to transform technological progress into broader economic and societal advancements across the population \citep{burton2023misq, simchi2020editor}.

However, the digital divide is often context dependent and can manifest differently across technologies \citep{haslberger2023no}. As technologies have advanced, research into digital disparities has expanded from internet access to various digital platforms and, more recently, artificial intelligence (AI), giving rise to a new subfield termed the ``AI divide'' \citep{mcelheran2024ai}. Compared to earlier ICTs, LLM-based generative AI exhibits distinctive properties, raising doubts about whether findings from prior ICT contexts can apply. First, its natural-language interface enables highly customizable inputs and produces outputs that are relatively easy to consume, which may help address users' information needs efficiently, particularly in long-tail domains \citep{xu2023chatgpt}. At the same time, system outputs are typically unstructured, making the assessment of model capabilities more salient and, in most settings, more difficult \citep{zhong2024agieval}. Moreover, system outputs may fail to align with users' intentions and exhibit hallucinations, further increasing noise in delivered information \citep{xu2024hallucination}. Given these unique properties, it is essential to examine how they translate into downstream disparities in adoption and, more broadly, to extend digital divide research to the context of generative AI (thereafter generative AI divide).

\subsection{Research Questions and Findings}

Building on previous research on the digital divide, we also conceptualize the generative AI divide in three layers \citep{wei2011conceptualizing, dewan2005digital}: (1) the \emph{access divide}, referring to inequalities in access to generative AI; (2) the \emph{capability divide}, concerning individual differences in ability to utilize generative AI once accessed; and (3) the \emph{outcome divide}, describing disparities in final outcomes, such as productivity and performance, which the preceding layers and other factors may influence. Furthermore, literature on IT and AI adoption has long emphasized the importance of individuals' perceptions during the adoption process (e.g., perceived usefulness and perceived ease of use), as perceived benefits influence adoption decisions \citep{jo2023analyzing, venkatesh2003user}. However, perceptions are dynamic and intangible, making it difficult to monitor \textit{how} and \textit{why} they change using data from surveys or experiments. Such dynamics are particularly salient in the generative AI context, where evaluating system capability itself is challenging and constantly evolving \citep{zhong2024agieval} and where alignment related issues introduce additional noise \citep{xu2024hallucination}.

Therefore, individuals' ability to update their perceptions of generative AI can substantially shape their adoption process. This view is consistent with the literature on AI literacy, which refers to the ability to understand, use, and evaluate AI systems \citep{annapureddy2025generative, ng2021conceptualizing}. Because AI literacy varies widely across individuals with different social characteristics \citep{wang2024artificial, celik2023exploring, robinson2020digital}, we aim to quantitatively measure and document disparities in individuals' ability to update their perceived utility through repeated interactions and other informational resources. We term this disparity the \textit{``learning divide,''} a novel aspect of the second layer \textit{capability divide}. The learning divide is consequential because heterogeneity in belief updating shapes adoption and usage decisions, thereby affecting the welfare individuals ultimately receive (i.e., the \textit{outcome divide}). For example, if an individual mistakenly concludes that the system is entirely useless after a single trial, they may stop using it and forgo opportunities to improve their work and quality of life through the technology. Conversely, if another individual forms an accurate understanding after initial use, they can use the system effectively and realize greater gains. To empirically examine the learning divide, we pose the following research question:

\textit{RQ1 (learning divide): How do individuals with different social characteristics update their perceived utility of generative AI at varying rates?}

Given that individuals' perceptions are unobservable and must be inferred from data, and their adoption decision-making is subject to informational uncertainty, structural learning models are ideally suited to address this problem \citep{wang2024identification, fang2022effects}. Therefore, to address the first research question, we develop a customized structural learning model to characterize individuals' dynamic learning and adoption processes of generative AI, allowing for heterogeneous learning speeds related to their social characteristics.

In parallel, individuals' disparate utility derived from generative AI upon usage has become the spotlight of recent studies in generative AI adoption, but now findings are mixed. On the one hand, earlier research on AI's impact on productivity suggests that workers with less prior expertise gain greater benefits from utilizing ChatGPT for specific tasks, potentially ameliorating existing inequalities \citep{noy2023experimental}. On the other hand, studies on AI adoption note that performance disparities have widened with the adoption of generative AI like ChatGPT \citep{mcelheran2024ai, haslberger2023no}. One of the potential explanations for the mixed findings is that most of these works are conducted in controlled \textit{experimental} settings where participants are required to use one of specific generative AI tools to complete \textit{predetermined} tasks. However, it remains underexplored how individuals perform in \textit{real-world} settings where they can use those general-purpose technologies for \textit{diverse} tasks. To supplement previous studies, we aim to uncover the differences in utility gains when individuals with various social characteristics use generative AI tools. We term this the \textit{``utility divide,''} and pose the following research question:

\textit{RQ2 (utility divide): How do individuals with different social characteristics gain utility differently from generative AI when they opt to use it?}

Consistent with research on digital divide employing behavioral experiments \citep{kohli2024digital, haslberger2023no, robinson2020digital}, our \textit{utility divide} evaluates disparities in utility per use, conditional on adoption, constituting a specific type of \textit{third-level outcome divide}. However, investigating this phenomenon in a real-world context is challenging due to the natural feedback loop in the adoption process. When individuals believe that a technology has high utility, they are more likely to adopt it. Upon using the generative AI tool, users update their beliefs based on their experiences (signals within the structural learning model), which in turn influence subsequent adoption decisions. Since individuals lack \textit{a priori} knowledge of the actual utility, it influences their adoption decisions only indirectly through signal realizations, which are also governed by their learning abilities. Therefore, recognizing that the utility divide and learning divide are intertwined, we incorporate utility heterogeneity into the structural learning framework to jointly model both disparities.

To empirically estimate our model and address the abovementioned research questions, we obtain a dataset from a leading data collection firm. The dataset comprises online footprints of thousands of panelists over a six-month period following the launch of ChatGPT, enabling us to observe individual usage patterns over time. Additionally, the company also provides social characteristics for each participant, allowing us to explore how social characteristics shape the proposed divides. After fitting our model to the dataset, we uncover significant \emph{learning} and \emph{utility divides}. Notably, individuals without college education, non-white individuals, and those with lower English literacy realize larger utility gains from ChatGPT, yet revise their beliefs about its utility more slowly. In contrast, males, younger individuals, and workers in occupations with greater exposure to LLMs both extract higher utility per use and update their perceived utility more quickly.

Although we have empirically validated both the \textit{learning divide} and the \textit{utility divide}, it remains unclear how these disparities influence individuals' lifetime usage and welfare dynamics, another form of \textit{outcome divides}. The \textit{utility divide} is, by definition, straightforward to interpret since it directly measures the utility gain per usage. However, the impact of the \textit{learning divide} is less evident, as it requires translating differences in learning speed into variations in adoption dynamics. Furthermore, it remains unclear how effective counterfactual policies might be in influencing these outcome divides through the learning process. Addressing this gap also aligns with the call for actionable insights in the literature on digital divide \citep{wang2024artificial, wei2011conceptualizing}. Therefore, we propose our final research question:

\textit{RQ3: How does the learning divide affect the outcome divide, and how to mitigate this outcome disparity?}

To address this research question, we leverage our estimated model parameters to simulate individuals' belief evolution trajectories over time. Through these simulations, we identify a phenomenon we term the \textit{``belief trap''}, in which users who initially form low perceptions of generative AI's utility may cease using it, thereby forgoing opportunities to update their beliefs through further interactions. This self-reinforcing cycle bears striking similarities to the ``poverty trap'' in development economics, where individuals or nations become locked in persistent poverty without external intervention \citep{banerjee2011poor}. Our analysis further shows that users with disadvantaged learning ability, as identified by our structural model, are substantially more likely to fall into this belief trap, constituting another manifestation of the \textit{outcome divide}. We also quantify the extent of news exposure or product improvement required to reduce belief trap risk and sustain long-run adoption, underscoring how difficult it can be to narrow these divides. Finally, our counterfactual simulations indicate that targeted training programs that increase exposure to generative AI can meaningfully lower trapping probabilities and reduce disparities, suggesting a feasible policy intervention to address this new form of digital divide.

\subsection{Theoretical Contributions}

This study makes several theoretical contributions. First, we present a unified framework for understanding IT adoption dynamics and the digital divide, particularly in scenarios characterized by information uncertainty and learning dynamics. Second, to empirically test this framework, our study leverages large-scale real-world usage data in which interaction with generative AI is voluntary, repeated, and consequential. This setting allows us to observe how beliefs about system capability evolve endogenously over time, how early noisy experiences shape subsequent adoption decisions, and how users self-select into or out of learning opportunities. As a result, we are able to uncover dynamic mechanisms that are mechanically ruled out in experimental designs with mandated exposure. Therefore, our paper presents a complementary approach to prior experimental studies that examine short-run productivity or learning effects under controlled exposure \citep{humlum2024adoption, noy2023experimental, haslberger2023no}. Third, our paper introduces two novel theoretical constructs, the \emph{learning divide} and the \emph{belief trap}, which are broadly applicable and offer new perspectives on technology adoption barriers. Fourth, with regard to the emerging literature on generative AI, we are among the first to document social characteristics-based disparities in both utility per use and learning ability and to quantify how these differences translate into adoption outcomes through a structural model. Finally, we map our empirical estimates into counterfactual policy evaluations, contributing evidence on the effectiveness of potential interventions to mitigate divides in generative AI adoption.

\subsection{Organization of the Paper}
The remainder of the paper is organized as follows. Section 2 reviews the related literature and positions our study within past research streams. Section 3 then develops the theoretical arguments about the generative AI divide. Section 4 describes the data sources, variable construction, and exploratory analyses. Section 5 presents the model, while Section 6 details the estimation procedures and identification strategy. Section 7 reports the estimation results and addresses RQ1 and RQ2. In Section 8, we conduct numerical experiments and evaluate counterfactual scenarios to address RQ3. Finally, Section 9 concludes with a broader discussion and outlines directions for future research.

\section{Related Literature}

Our paper contributes to three streams of literature: digital divide, AI adoption, and dynamic learning processes. In each subsection, we review findings from previous literature, identify the research gap, and highlight how our study extends the literature. 

\subsection{Digital Divide}
The digital divide refers to the disparities in access to, capability with, and effective use of information and communication technologies (ICT) among individuals from diverse backgrounds \citep{vassilakopoulou2023bridging, oecd2001understanding}. As a critical societal issue, it has been extensively investigated (see reviews by \cite{vassilakopoulou2023bridging, lythreatis2022digital, scheerder2017determinants, dewan2005digital}), with key references summarized in Appendix Table \ref{tab:digitaldivide}. Reflecting its complex, multidimensional nature, the digital divide manifests at three distinct levels \citep{wei2011conceptualizing, dewan2005digital}: (1) the first-level \textit{access} divide, concerning inequalities in access to ICT; (2) the second-level \textit{capability} divide, addressing disparities in the ability to effectively use ICT among those with access; and (3) the third-level \textit{outcome} divide, describing differences in outcomes due to variations in access, usage patterns, and other related factors \citep{wei2011conceptualizing}. 

Prior studies have revealed a series of key factors contributing to the digital divide across multiple levels and dimensions, including social characteristics \citep{venkatesh2014understanding, fairlie2004race, jackson2001racial}; personal attributes and skills (e.g., motivation and personality traits) \citep{celik2023exploring, hsieh2011addressing, hargittai2002second}; social and environmental influences (e.g., social support networks and geographical location) \citep{kohli2024digital, hargittai2003informed}; and technical factors (e.g., platform affordance) \citep{wang2024artificial, yang2023beyond}. Among them, social characteristics have been identified as central determinants of the digital divide \citep{venkatesh2014understanding, fairlie2004race, jackson2001racial}. The social characteristics, including education \citep{hidalgo2020digital, hsieh2008understanding}, age \citep{jackson2001racial}, gender \citep{van2003digital, hargittai2003informed}, race \citep{fairlie2004race}, language proficiency \citep{bassignana2025ai, hargittai2003informed, hargittai2002second}, and occupational position and job roles \citep{mcelheran2024ai, venkatesh2014understanding}, have been shown to shape individuals’ access to ICT, their ability to use it effectively, and the outcomes they derive across contexts and over time \citep{celik2023exploring, scheerder2017determinants}. Importantly, prior literature emphasizes digital literacy as a key mechanism through which social characteristics translate into differential ICT usage and outcomes, mediating the relationship between social characteristics and realized benefits \citep{scheerder2017determinants, hargittai2002second}. This mechanism is especially salient in the context of generative AI, where effective use depends on formulating appropriate prompts, interpreting probabilistic and unstructured outputs, and evaluating system capability under uncertainty.

In addition, the manifestation of the digital divide exhibits substantial variation across technologies and platforms \citep{yang2023beyond, burtch2019investigating}, and the effects of social characteristics on adoption and usage patterns may differ depending on the specific technology or context \citep{haslberger2023no}. As such, the digital divide is a context-specific concept, which makes it necessary to specify the type of ICT under investigation \citep{vassilakopoulou2023bridging}. As a result, the discussion about the digital divide has evolved alongside the progression of digital technologies. Early studies focus on the unequal access to and use of personal computers \citep{autor1998computing}, while later researchers expanded to include Internet access and use \citep{van2003digital, hargittai2002second}. Then, the focus shifted towards the use of digital platforms, such as e-government platforms \citep{venkatesh2014understanding}, medical crowdfunding platforms \citep{burtch2019investigating}, and online rental housing platforms \citep{boeing2020online}. Most recently, the diffusion of AI has given rise to a new dimension of inequality, the AI divide \citep{mcelheran2024ai}. Understanding this divide in the context of generative AI is particularly urgent, as it dramatically differs from earlier ICTs in its probabilistic, adaptive, and interactive nature. Therefore, it is meaningful to explore what new forms of inequality these distinctive technical properties generate and how social characteristics shape capability and outcome divides in the context of generative AI.

\subsection{AI Adoption} 

Our paper contributes to the emerging literature on AI adoption, summarized in Appendix Table \ref{tab:aiadoption}. Prior research has focused on the \textit{antecedents} and \textit{consequences} of adopting traditional AI technologies \citep{wang2025artificial, mcelheran2024ai, chen2024does, alekseeva2020ai}. In exploring the \textit{antecedents}, studies have adopted a behavioral perspective to investigate how technological, organizational, and environmental factors influence AI adoption decisions \citep{jo2023analyzing, prasad2023towards}. The majority of studies employ behavioral methods, including surveys, interviews, and controlled experiments, to deliberate on users' perceptions of AI (e.g., perceived ease of use and perceived usefulness) and how these perceptions affect adoption incentives \citep{gupta2024adoption, choudhury2023investigating}. Regarding the \textit{consequences}, it is well established that although AI adoption produces economic benefits, these benefits may be unevenly distributed. For instance, \citet{agrawal2024artificial} suggests that unequal access to and capabilities in utilizing non-generative AI technologies can exacerbate disparities among firms in terms of organizational performance and market dynamics. This issue is concerning, as it may lead to a widening gap between firms able to harness the benefits of AI and those that cannot, resulting in significant disparities in innovation, productivity, and market competitiveness \citep{gans2023artificial}.

Our paper models generative AI adoption using real-world data and thus complements experimental evidence and survey studies on AI adoption in two ways. First, prior experimental studies on generative AI provide important causal evidence on short-run productivity and learning effects under controlled exposure, involving fixed tasks, mandated use, and a limited interaction time window \citep{humlum2024adoption, noy2023experimental}. In contrast, our study examines repeated, voluntary use of generative AI over an extended observational window, where users can freely experiment, reduce usage, or abandon generative AI altogether. It allows us to uncover self-reinforcing mechanisms such as belief traps, in which early negative or ambiguous experiences lead users to prematurely exit and thereby forgo further belief updating, dynamics that cannot be detected in short-horizon or forced-use experimental designs. Second, the real-world setting allows us to disentangle the learning and utility divides. In experimental settings, learning and utility divides are intertwined because usage intensity, learning opportunities, and outcomes are constrained by design rather than users' choice. In contrast, in real-world settings, learning speed, per-use utility, and adoption decisions are determined by users’ own choices under uncertainty rather than by experimental protocols. By combining rich behavioral data with a structural Bayesian learning framework, we are able to reveal how inequality emerges dynamically when learning unfolds over time and adoption is voluntary.

\subsection{Learning and Decision-Making under Uncertainty}

When confronted with \textit{ex ante} information uncertainty or incompleteness, decision-makers often rely on learning processes to resolve uncertainty and make more informed decisions. Structural Bayesian learning models have thus been proposed to characterize these dynamic learning processes \citep{wang2024identification, hu2022human, erdem1996decision}, and our model follows this stream of literature.

These models have been most commonly applied in the context of product attribute assessment, particularly in evaluating product quality \citep{erdem1996decision}. Consumers iteratively update their beliefs regarding product attributes by utilizing various information signals, including direct experience, advertising exposure, and online reviews \citep{lin2015learning, erdem2008dynamic}. The versatility of these models extends to more complex domains, such as pharmaceutical efficacy evaluation \citep{ching2020structural}, consumption of addictive products \citep{chen2020dynamic}, and financial investment decisions \citep{zhou2021disclosure}. Moreover, Bayesian learning models have been adapted to investigate diverse user behaviors, encompassing phenomena such as opinion polarization \citep{lu2022microblogging}, application usage patterns \citep{zheng2020optimizing, zhang2020learning}, rating behaviors \citep{ho2017disconfirmation}, ideation processes \citep{huang2014crowdsourcing}, and content generation dynamics \citep{ghose2011empirical}.

Our research context where individuals learning about generative AI through repeated interactions that are subject to randomness, presents an ideal setting for applying structural learning models. The inherent variability in AI system outputs creates uncertainty that users must resolve through experience, making the Bayesian learning framework particularly relevant. To our knowledge, this study is among the first to utilize a learning model to characterize heterogeneous AI adoption patterns and to quantify the learning process as a mechanism contributing to the AI divide. Furthermore, given the ongoing evolution of AI technologies, our proposed framework offers potential for longitudinal monitoring of dynamic adoption trends.

\section{Theoretical Development}
\begin{figure}[H]
    \centering
    \begin{minipage}{1.0\textwidth}
        \centering
        \includegraphics[width=\textwidth]{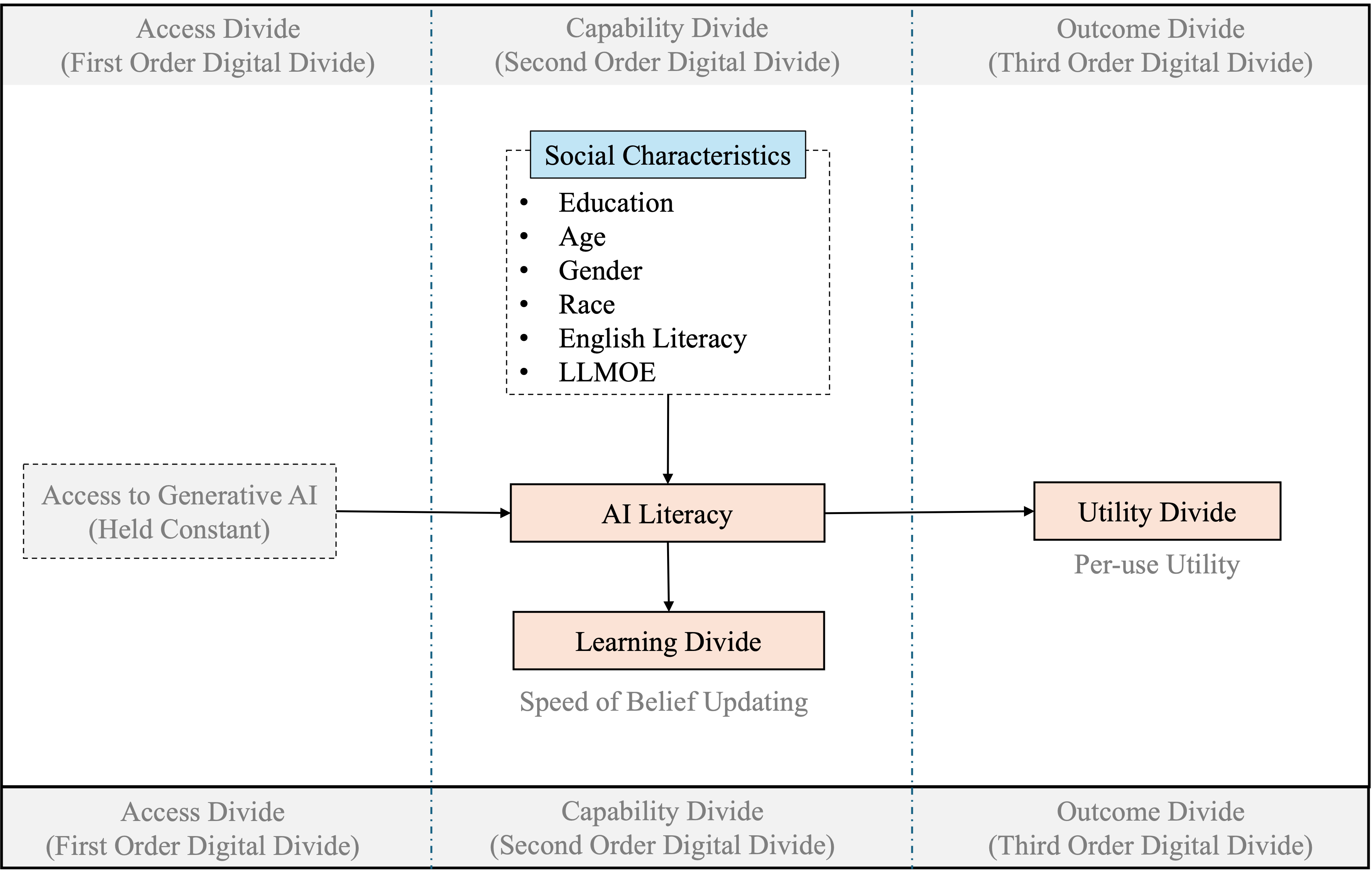}
    \end{minipage}

    \caption{Conceptual Framework of Generative AI Divide}
    \label{fig:theory}
\end{figure}

In this section, we develop a theoretical framework for generative AI divide and explain why the underlying mechanisms may differ from the divides in traditional ICT adoption. First, we highlight key technical features of generative AI that can generate distinctive behavioral patterns. Then, we theorize how social characteristics shape generative AI outcomes through AI literacy, giving rise to learning and utility divides at the capability and outcome layers even when access is held constant. Figure~\ref{fig:theory} illustrates how our study is situated in the literature on the digital divide and how these mechanisms unfold in the generative AI setting. 

\subsection{Key Technical Distinctions Between Generative AI and ICTs}

Generative AI models operate on a generative and probabilistic architecture. It produces responses via probabilistic token generation: when a user queries, the system generates outputs by sampling from a learned conditional distribution given the prompt \citep{morris2023levels, kojima2022large}. In contrast, conventional ICTs, such as personal computers, software applications, and search engines, are relatively more predictable. They help users locate and interact with information and services that exist within or outside the system \citep{dewan2005digital, venkatesh2003user}. These technological properties introduce fundamental distinctions between generative AI and prior ICTs.

First, generative AI can generate responses that adhere to user instructions (prompts). Most generative AI systems undergo instruction tuning, which enables them to follow specific instructions and even examples embedded in the prompt (in-context learning) to shape their outputs accordingly \citep{liu2025deepseek, zhao20247b, ouyang2022training}. Consequently, well-crafted and individualized prompts can return personalized and niche outputs that may be difficult to surface through conventional keyword-based search queries \citep{xu2023chatgpt}. Second, generative AI produces unstructured outputs through a natural-language interface, which complicates the assessment of output quality and system capability \citep{zhong2024agieval}. Many prior ICTs can be evaluated using standard metrics such as accuracy and speed and easily verified. However, the output quality of generative AI is often less transparent, and supporting evidence for responses may be absent or difficult to verify, increasing the need for careful evaluation \citep{ma2024exploring, choudhury2023investigating}. Third, generative AI introduces greater uncertainty and signal noise. Because generation prioritizes statistical plausibility rather than factual fidelity, responses can be fluent yet incorrect or fabricated (``hallucinations'') \citep{xu2024hallucination}. Consequently, user feedback is noisier than in traditional ICT systems. This heightened uncertainty alters how users evaluate and update beliefs about system capability over time, making learning more important yet fragile \citep{yan2024promises}.

\subsection{From Technical Distinctions to Generative AI Divide}

Generative AI introduces inequality because variation in AI literacy shapes how users learn from and extract value from the system over time. In our context, AI literacy manifests as the ability to formulate prompts effectively, interpret outputs accurately, and adapt usage strategies under uncertainty. First, at the capability layer, the generative nature of generative AI produces unstructured outputs, and the quality of responses to open-ended questions cannot be evaluated by accuracy alone, giving rise to a learning divide. Users have to infer system capability through repeated interactions. Those with higher AI literacy are better able to integrate information across interactions and update their beliefs, enabling faster convergence toward accurate assessments of system capability \citep{ng2021conceptualizing}. Second, at the outcome layer, the instruction-following and in-context learning capabilities of generative AI create a utility divide through prompt-dependent value creation. Because the output quality depends on how users formulate prompts and frame tasks, users with higher AI literacy can elicit more relevant and accurate responses, thereby obtaining greater utility per use \citep{annapureddy2025generative}. Third, the stochastic and uncertain nature of generative AI amplifies heterogeneity at the capability and outcome layers. Early interactions generate noisy and potentially misleading signals, which increase dispersion in belief updating across users and heighten the risk of mislearning or premature abandonment among users with lower AI literacy \citep{yan2024promises}. Meanwhile, uncertainty increases variance in realized per-use utility, as identical prompts may yield different outcomes across interactions. These dynamics generate self-reinforcing belief traps that operate within the learning and utility divides.

In our context, AI literacy is a generative AI specific extension of digital literacy that is socially produced \citep{celik2023exploring}. Consistent with research on the digital divide, AI literacy is shaped by socio-demographic characteristics (e.g., age, gender, and race) and socio-economic attributes (e.g., education, English literacy, and occupation) \citep{wang2024artificial,celik2023exploring,robinson2020digital,scheerder2017determinants,dewan2005digital,van2003digital}. Variation in these social characteristics might translate into systematic differences in users' ability to interact with, interpret, and evaluate generative AI systems, which in turn generate learning and utility divides \citep{vassilakopoulou2023bridging,lythreatis2022digital}. Specifically, users who are better educated, younger, male, white, more proficient in English, and employed in occupations with greater exposure to large language models are more likely to possess higher AI literacy \citep{bassignana2025ai, mcelheran2024ai,fairlie2004race, jackson2001racial}, enabling faster learning about system capabilities and higher per-use utility from interactions with generative AI \citep{annapureddy2025generative, celik2023exploring}. Appendix \ref{appendix:theory} elaborates more on the rationales.

\section{Data, Variable, and Exploratory Analysis}
To empirically test our research questions, we use a real-world dataset from a clickstream tracking company, which offers several advantages \citep{chiou2022restrictions,chesnes2017banning,zheng2012business}. First, it records all URLs visited by each participant over an extended period, enabling us to observe the temporal dynamics of generative AI adoption and use. Second, in addition to clickstream records, it includes measures of participants' social characteristics, allowing us to link these characteristics to usage patterns and empirically examine the AI divide. Third, relative to laboratory settings, this real-world data measurement enables us to reveal how inequality emerges dynamically in voluntary adoption when learning unfolds over time and uncover self-reinforcing mechanisms such as belief traps.

In this section, we describe the data source and collection procedures. We then assess the representativeness of the sample. 
Finally, we elaborate on the variable construction and the rationale underlying our measures to support the structural estimation.

\subsection{Data Source}

The data tracking company (thereafter the company) advertises for and recruits participants to contribute data in exchange for compensation. Those participants are referred to as ``panelists.'' To enroll, panelists complete a registration form providing social characteristics information, such as gender and age. The company verifies this information prior to approval. Once approved, panelists install a tracking application on their primary devices. The application then continuously records their clickstream activity. The company conducts ongoing targeted outreach and recruitment across demographic groups to ensure that the panel remains broadly representative of the U.S.\ population.

To study the dynamics of ChatGPT adoption, we obtain a dataset covering November 30, 2022 (the initial public release of ChatGPT) through May 31, 2023. This six-month window allows us to observe how individuals discover, experiment with, and potentially incorporate ChatGPT into their routines. Our sample includes 11,752 users who were active for at least 60 days during the observation period and have their demographic profiles verified. Among them, 1,386 users had used ChatGPT at least once. Within this cohort, 1,344 users interacted with ChatGPT at least twice, providing the repeated observations necessary to identify choice dynamics and learning \citep{ho2017disconfirmation}. In addition, we obtain the November data for the full sample to support the construction of pre-treatment behavior-based variables, as described in the next subsection.

\subsection{Variable Construction and Exploration}

To examine the potential presence of digital divide in ChatGPT adoption, we link individual differences in ChatGPT usage to their social characteristics. Based on our theoretical development (Section 3), we examine how education, age, gender, race, occupation, and English literacy shape the AI adoption process dynamically. Education has been identified as one of the most robust predictors of digital disparities \citep{hidalgo2020digital, hsieh2011addressing, hsieh2008understanding}. In addition, digital literacy has been shown to be strongly associated with age, with younger cohorts demonstrating greater proficiency about technologies \citep{wang2024artificial, jackson2001racial}. Prior studies further document significant roles of gender disparities \citep{van2003digital, jackson2001racial} and racial differences \citep{boeing2020online, fairlie2004race} in the digital divide. Occupation plays a crucial role in shaping digital skills and usage patterns \citep{hidalgo2020digital, venkatesh2014understanding}, and, more relevant to generative AI adoption, the exposure to LLM-relevant tasks can substantially influence individuals' adoption behavior \citep{eloundou2024gpts, liu2023generate}. Lastly, because LLM performance varies across languages and often favors English, we also assess the impact of users' English literacy \citep{martinez2025llms, qin2025survey}.

We then construct a set of variables to operationalize these factors. Education is captured by $College$, a binary variable equal to 1 if a user has some college education and 0 otherwise. Age is grouped into three categories (low, middle, and high) based on the sample tertile cutoffs (35 and 47 years), with the middle and high categories represented by the dummy variables $AgeMid$ and $AgeHigh$, respectively. Gender is measured by $Male$, a binary variable equal to 1 for male users and 0 otherwise. Racial background is represented by $White$, a dummy variable indicating whether the user identifies as white. To measure occupational exposure to LLMs (LLMOE), we construct the $LLMOE$ score from \citet{eloundou2024gpts}: we match each user's occupation to the occupations reported in \citet{eloundou2024gpts} via semantic similarity and then assign the $LLMOE$ value of the most similar occupation. Finally, $English$ proxies for English literacy and is measured as the share of English language queries among all search queries made prior to ChatGPT's launch. The summary statistics are reported in Appendix~\ref{app:summary_stats}. To assess the representativeness of our dataset, we compare the distributions of social characteristics in our sample with those in the U.S. population. Appendix~\ref{app:comparison_with_population} presents the comparison. Overall, there is no significant difference across social characteristics of the two distributions, suggesting that our sample is reasonably representative. As reported in Appendix~\ref{app:corr_matrix}, we also test the pairwise correlation of each pair of social characteristic variables, and all correlations are smaller than 0.2, which indicates that multicollinearity is unlikely to be a concern in our setting.

We next present exploratory evidence of potential divides in generative AI adoption and usage. First, Appendix~\ref{app:variable_distribution} compares the demographic distributions of the full sample with those of ChatGPT users. The comparisons suggest that adoption of ChatGPT is more likely among younger individuals, men, non-white individuals, those with some college education, those with lower English literacy, and those with higher LLMOE. Second, our reduced form analyses reinforce these patterns. Specifically, within the repeat ChatGPT user subset, we regress the average daily ChatGPT usage frequency, $GPTFrequency$, on the same set of social characteristics. As Appendix~\ref{app:reduce_form} reports, the groups highlighted in the graphical evidence (younger, male, with some college education, with lower English literacy, and with higher LLMOE) also use ChatGPT more frequently. Taken together, these results suggest the presence of a generative AI divide, which further underscores the importance of our research questions.

\subsection{Variable Construction for the Bayesian Learning Model}
\label{subsec: variable}

To further disentangle utility heterogeneities and learning ability heterogeneities as underlying mechanisms of the adoption patterns, we develop a structural model that identifies the dynamic learning process while accounting for utility differences.

Modeling generative AI adoption requires an outside option against which to benchmark users' decisions. We use search engine activity as this option, which is directly observed in our dataset. Because ChatGPT is trained on a vast corpus to answer a wide range of questions \citep{morris2023levels}, it can serve as an alternative tool for accessing information. Our choice is also backed up by a growing literature that compares user search behavior in ChatGPT versus traditional search engines across different scenarios \citep{sun2024trusting, karunaratne2023new, xu2023chatgpt}. Leveraging users' time-varying usage of ChatGPT relative to search engines, we can trace perception dynamics and disentangle the underlying mechanisms. 

We construct the variables used in the Bayesian learning model in several steps. From the full sample, we retain 1,344 users who used ChatGPT at least twice, and the repeated observations allow us to observe belief updating \citep{ho2017disconfirmation}. We then aggregate each user's clickstream data at the daily level, yielding up to at most 183 observations per user, with each observation corresponding to a single day. Because our focus is on the dynamic learning process following the initial exposure to ChatGPT, we exclude records prior to a user's first exposure to ChatGPT, whether that exposure occurs through consuming related news or through direct use. After preprocessing, the final dataset contains 134,022 observations from 1,344 users. Table~\ref{tab:summary_statistics} reports summary statistics for all variables.

\begin{table}[!h] \centering \small
  \caption{Summary Statistics (Main Model)} 
  \label{tab:summary_statistics} 
\begin{tabular}{@{\extracolsep{5pt}}llrrrrrr} 
\toprule
Statistic & \multicolumn{1}{c}{Variable} & \multicolumn{1}{c}{N} & \multicolumn{1}{c}{Mean} & \multicolumn{1}{c}{St. Dev.} & \multicolumn{1}{c}{Max} & \multicolumn{1}{c}{Min} & \multicolumn{1}{c}{Median} \\ 
\midrule
\textbf{Observation-level Variables}   \\
\textit{Total Activities(Search + ChatGPT)} & $W_{itd}$ & 134,022 & 16.49 & 40.89 & 2,658 & 0 & 3 \\ 
\textit{ChatGPT Times} & $W_{its}$ & 134,022 & 1.18 & 10.49 & 2,651 & 0 & 0 \\ 
\textit{Search Times} & $W_{itd} - W_{its}$ & 134,022 & 15.31 &  30.98 & 1,297 & 0 & 2 \\ 
\textit{News Exposure} & $W_{itn}$& 134,022 & 0.01 & 0.40 & 130 & 0 & 0 \\ 
\textbf{Individual-level Variables}   \\
\textit{College Education} & $College_i$ & 1,344 & 0.56 & 0.50 & 1 & 0 & 1 \\
\textit{Age - High Group(>47)} & $AgeHigh_i$ & 1,344 & 0.23 & 0.42 & 1 & 0 & 0 \\
\textit{Age - Middle Group(35 \textasciitilde{} 47)} & $AgeMid_i$ & 1,344 & 0.29 & 0.45 & 1 & 0 & 0 \\
\textit{Male} & $Male_i$ & 1,344 & 0.60 & 0.49 & 1 & 0 & 1 \\
\textit{White} & $White_i$ & 1,344 & 0.65 & 0.48 & 1 & 0 & 1 \\
\textit{LLM Occupational Exposure} & $LLMOE_i$ & 1,344 & 0.12 & 0.10 & 0.79 & 0 & 0.11 \\
\textit{English Literacy} & $English_i$ & 1,344 & 0.79 & 0.15 & 1 & 0 & 0.83 \\
\bottomrule
\end{tabular} 
\end{table}  

Each time a user $i$ accesses online information resources, they may use either ChatGPT or traditional search engines\footnote{For traditional search engines, we focus on Google and Microsoft Bing, which together account for the majority of U.S.\ search engine market share.}. Let $W_{itd}$ denote the total number of information-seeking activities for user $i$ on day $t$, and let $W_{its}$ represent the number of ChatGPT activities on that day\footnote{One ChatGPT activity is defined as a user visiting the ChatGPT website (i.e., opening a ChatGPT page). We treat an activity as ending when the user leaves the ChatGPT page to visit other pages, or when the user opens a new ChatGPT chat that changes the URL.}. The remaining activities, $W_{itd}-W_{its}$, correspond to usage of traditional search engines. Moreover, to account for other online information sources that may influence beliefs \citep{wang2024identification}, we construct $W_{itn}$ to capture the number of times user $i$ clicks on ChatGPT-related news articles on day $t$\footnote{The set of news outlets was compiled from the following website as of April 2024: \url{https://pressgazette.co.uk/media-audience-and-business-data/media_metrics/most-popular-websites-news-us-monthly-3/}.}. Finally, consistent with our exploratory analysis, we include $College_i$, $AgeHigh_i$, $AgeMid_i$, $Male_i$, $White_i$, $LLMOE_i$, and $English_i$ as sources of heterogeneity underlying the generative AI divide in the Bayesian learning model. 

\section{Model}
\label{sec: model}

Motivated by our theoretical inquiry and exploratory analyses, we develop a Bayesian learning model to investigate the dynamics of adoption behaviors. Whenever a user would like to access resources from the Internet, she must decide whether to start the process with ChatGPT or traditional search engines. This decision is based on the perceived utility derived from each option, which evolves through a learning process influenced by the user's prior experiences. Consequently, the final choice is determined through utility maximization. The utility function, learning process, and decision-making procedure will be elaborated upon in the following subsections.

\subsection{Utility Function}
\label{subsec: utility}

When user $i$ has a general query requiring online resources at time $t$, they may either utilize a traditional search engine or opt for the newly launched ChatGPT, which is trained on a vast corpus of existing information \citep{xu2023chatgpt}. To rationalize the decision-making process between adopting ChatGPT versus the traditional search engine (considered as the outside option in this technology adoption scenario), we denote $u_{it1}$ as the utility that user $i$ derives from choosing ChatGPT at time $t$, and $u_{it0}$ as the utility derived from choosing the outside option.

Given that the traditional search engine has been active for a long period, we use it as the baseline and assume that its utility $u_{it0}$ can be decomposed into a constant term $c_i$ and a stochastic error term $\varepsilon_{it0}$, in line with the Bayesian learning literature \citep{zheng2020optimizing}:
\begin{align}
    u_{it0} = c_i + \varepsilon_{it0}
\end{align}

Similarly, we decompose the utility of choosing ChatGPT into a representative utility term, which can be explained by observable factors, and a stochastic error term $\varepsilon_{it1}$.
\begin{align}
    u_{it1} = v_{i} + \varepsilon_{it1} 
\end{align}

As introduced in RQ2, users with different social characteristics may derive varying levels of utility from using ChatGPT. Therefore, we model each individual's attributes as contributing linearly to her representative utility, allowing us to characterize the potential \textit{utility divide}.
\small
\begin{align}
v_{i} = &\ \alpha_{0i} + \alpha_1 \cdot College_i + \alpha_2 \cdot AgeHigh_i + \alpha_3 \cdot AgeMid_i + \alpha_4 \cdot Male_i \notag\\
&\ + \alpha_5 \cdot White_i + \alpha_6 \cdot LLMOE_i + \alpha_7 \cdot English_i
\end{align}
\normalsize

Note that in other discrete choice settings \citep{train2009discrete}, only differences in utility are relevant; therefore, only one constant term is needed in the utility functions of the two alternatives to capture the baseline relative differences between ChatGPT and the traditional search engine. However, in our learning model, where we normalize the distribution of individuals' initial beliefs, both constants can be identified, as discussed in Section \ref{sec:est_identification}. Additionally, the parameters $(\alpha_1, \ldots, \alpha_7)$ also capture differences in relative utility instead of absolute utility. For instance, a positive $\alpha_1$ would indicate that users with higher education tend to derive greater relative gains from using ChatGPT compared to traditional search engines.

Since ChatGPT has been launched only recently, individuals may not initially be fully certain about ChatGPT's utility $u_{it1}$. Therefore, \textit{before} a user $i$ decides whether to use ChatGPT to fulfill her query, she first forms an expectation of the utility based on the information available up to period $t$ (hereinafter referred to as ``perceived utility''), and then choose the alternative that maximizes her expected utility. The perception formation process follows the Bayesian learning framework \citep{wang2024identification}, and will be elucidated in Section \ref{subsec: learning}.

Regarding the stochastic error terms in both utility functions ($\varepsilon_{it0}$ and $ \varepsilon_{it1}$), they represent utility shocks that influence individuals' decisions between the two alternatives but are unobserved by researchers. Consistent with the discrete choice literature \citep{train2009discrete}, we assume that these error terms follow a Type-I extreme value distribution, allowing the expected utility to be linked to the choice probability in a logit functional form.

\subsection{Individuals' Learning Process}
\label{subsec: learning}

As outlined in the previous section, individuals' choices of tools are influenced by their perceived utility of each option \textit{prior} to making a decision. This also aligns with the technology adoption literature, which emphasizes the importance of perception in the adoption of newly launched technologies \citep{ma2024exploring, venkatesh2003user}. Given that stochastic shocks ($\varepsilon_{it0}$ and $ \varepsilon_{it1}$) affecting consumers' decisions are known to consumers by definition \citep{zheng2020optimizing, train2009discrete}, the perceived utility can be represented as follows.
\begin{align}
    & E[u_{it0} | I_{it}] = E[c_i | I_{it}] + \varepsilon_{it0} \\
    & E[u_{it1} | I_{it}] = E[v_i | I_{it}] + \varepsilon_{it1}
\end{align}

Here $I_{it}$ represents all the information available to the user $i$ until period $t$ that can help them form perceptions of the utility of the two options for themselves.

Since traditional search engines have been available for an extended period, individuals are already familiar with them, and their beliefs are assumed to have stabilized \textit{prior} to the observation period. Consequently, similar to other Bayesian learning models \citep{zheng2020optimizing}, the perceived representative utility of the outside option $E[c_i | I_{it}]$ is assumed to be constant for consumers \textit{a priori} and is \textit{not} influenced by new information.
\begin{equation}
E[u_{it0} | I_{it}] = E[c_i | I_{it}] + \varepsilon_{it0} = c_i + \varepsilon_{it0}
\end{equation}

In contrast, the utility of the newly introduced ChatGPT is less certain to users, resulting in a dynamic update of their belief $E[u_{it1} | I_{it}]$ over time as they gather more information and experience with the tool. Following the Bayesian learning literature \citep{wang2024identification, erdem1996decision}, we assume that prior to any direct experience with ChatGPT, individuals possess some prior information regarding the representative utility of using ChatGPT, which is modeled as following a normal distribution $N(v_{0}, \sigma_0^2)$. The individual perceives that the actual representative utility $v_i$ is drawn from this distribution, where the mean of this distribution represents the mean perception $v_{0}$. $v_{0}$ may differ from the actual value $v_i$, which can reflect the initial skepticism about ChatGPT's representative utility. Users who are more skeptical may begin with a lower $v_{0}$, leading to a larger gap with the true representative utility $v_{0}$ and requiring a longer belief-updating process to converge toward the true representative utility. The variance $\sigma_0^2$, on the other hand, reflects the level of uncertainty in individuals' perceptions: all else being equal, a larger variance indicates greater uncertainty about their expected representative utility, making individuals more influenced by incoming information. Therefore, at period $0$, the individual's mean expectation $E[v_i | I_{i0}]$ and the variance of such mean expectation $\sigma_{i0}^2$ can be expressed as:
\begin{align}
    & E[v_i | I_{i0}] = v_{0} \nonumber \\
    & \sigma_{i0}^2 = E((v_i - v_0)^2 | I_{i0}) = \sigma_0^2
\end{align}

When users actually use ChatGPT, they gain first-hand experience regarding the utility derived from ChatGPT, thereby updating their beliefs about $v_i$ and the associated uncertainty. Consistent with the Bayesian learning framework, we denote the information obtained from such experiences as a signal $s_{itk}$, where individual $i$ uses ChatGPT at the $k$-th time during period $t$ \citep{erdem1996decision}. These signals are informative but inherently noisy, due to either variability in the actual representative utility or users' subjective perception processes \citep{zheng2020optimizing}. To maintain a conjugate update process \citep{van2021bayesian, degroot2005optimal}, we leverage the properties of normal distributions and assume that the signal follows a normal distribution \citep{wang2024identification}: 
\begin{align}
    s_{itk} \sim N(v_{i}, \sigma_{s_i}^2)
\end{align}

Consistent with prior Bayesian learning studies and the concept of rational expectations, the mean of the signal is assumed to equal the true representative utility \citep{zhang2020learning, huang2014crowdsourcing}. Regarding the signal variance, similar to any Bayesian updating process, it influences the relative weight given to prior beliefs versus the incoming signal when constructing the posterior belief, thereby reflecting individuals' learning speed in converging towards the actual representative utility \citep{ho2017disconfirmation, erdem1996decision}. As introduced by our RQ1 (learning divide), individuals with different backgrounds may exhibit heterogeneous learning abilities. To explore such potential disparities, we link observable attributes to the variance as specified in Equation \ref{eq:signal_hetero}. Specifically, within this equation, we apply an exponential transformation to a linear combination of attributes to ensure that the variance $\sigma_{s_i}^2$ remains positive and covers the entire positive domain.
\small
\begin{align}
\label{eq:signal_hetero}
\sigma_{s_i}^2 = &\ \exp(\gamma_{0i} + \gamma_1 \cdot College_i + \gamma_2 \cdot AgeHigh_i + \gamma_3 \cdot AgeMid_i+ \gamma_4 \cdot Male_i \notag\\
&\ + \gamma_5 \cdot White_i + \gamma_6 \cdot LLMOE_i + \gamma_7 \cdot English_i)
\end{align}
\normalsize

Furthermore, to control for other information sources that might influence individuals' perceptions, we incorporate news article access based on clickstream data as an additional source of signals. Specifically, at period $t$, when user $i$ clicks on the $j$-th news article that mentions ChatGPT, we model this signal as being drawn from a normal distribution $n_{itj} \sim N(v_{i}, \sigma_n^2)$.

We now further illustrate the perception updating process based on signals. To improve estimation efficiency, we aggregate data and analyze users' learning processes on a daily basis \citep{zhou2022learning}. At the beginning of day $(t-1)$, an individual's perceived representative utility of ChatGPT is characterized by a normal distribution with mean $E[v_i | I_{i(t-1)}]$ and variance $\sigma_{i(t-1)}^2$. Throughout day $(t-1)$, user $i$ interacts with ChatGPT for $W_{i(t-1)s}$ times and accesses $W_{i(t-1)n}$ related news articles. Using these signals, the user updates their belief according to the Bayesian rule \citep{degroot2005optimal, erdem1996decision}, forming a new belief at the end of the day that the actual representative utility is drawn from the distribution $N(E[v_i | I_{it}], \sigma_{it}^2)$ where
\begin{align}
E[v_i | I_{it}] &= \frac{\frac{E[v_i | I_{i(t-1)}]}{\sigma_{i(t-1)}^2} + \sum_{k=1}^{W_{i(t-1)s}}\frac{s_{i(t-1)k}}{\sigma_{s_i}^2} + \sum_{j=1}^{W_{i(t-1)n}}\frac{n_{i(t-1)j}}{\sigma_n^2}} {{\frac{1}{\sigma_{i(t-1)}^2} + \frac{W_{i(t-1)s}}{\sigma_{s_i}^2}} + \frac{W_{i(t-1)n}}{\sigma_n^2}} \\
\sigma_{it}^2 &= \frac{1}{\frac{1}{\sigma_{i(t-1)}^2} + \frac{W_{i(t-1)s}}{\sigma_{s_i}^2} + \frac{W_{i(t-1)n}}{\sigma_n^2}}
\end{align}

As additional signals are received, the perceived representative utility $E[v_i | I_{it}]$ gradually converges towards the true representative utility $v_i$, albeit with some noise, while the associated uncertainty $\sigma_{it}^2$ progressively diminishes towards zero.

Finally, according to the Bayesian updating equation, a larger signal variance implies a slower updating process, as the weight assigned to the noisy but unbiased signal is inversely related to the signal variance \citep{zheng2020optimizing}. For instance, a negative $\gamma_1$ would imply that, all else being equal, individuals with higher education are able to converge their perceptions $E[v_i | I_{it}]$ to the actual value $v_i$ more rapidly after receiving the same number of signals compared to those with lower education.

\subsection{Unobserved Heterogeneity and Likelihood Function}

In this binary choice scenario, each individual selects the option that maximizes their expected utility by comparing $E[u_{it1} | I_{it}]$ and $E[u_{it0} | I_{it}]$. Assuming that the unobserved utility shock follows a Type-1 extreme value distribution, the difference between these utilities is characterized by a logistic distribution \citep{train2009discrete}. This assumption enables a closed-form expression for the choice probability of selecting ChatGPT, based on the representative perceived utilities ($E[v_i | I_{it}]$ and $c_i$), using the logit model:
\begin{align}
  Pr(d_{itw} = 1) &= \frac{\exp(E[v_i | I_{it}])}
  {\exp(E[v_i | I_{it}]) + \exp(c_i)}
\end{align}

Here, $d_{itw}$ represents whether individual $i$ chooses ChatGPT at the $w$-th decision point within period $t$. Given that there are $W_{itd}$ decision opportunities in period $t$, the likelihood of observing the individual choosing ChatGPT for $W_{its}$ occasions and opting for search engines for the remaining $(W_{itd} - W_{its})$ occasions can be expressed as:
\begin{align}
\label{equ:lik_nolatent}
  l_{it}(W_{itd}, W_{its} \mid \bm{\theta}) &=  \left( \frac{\exp(E[v_i | I_{it}])}
  {\exp(E[v_i | I_{it}]) + \exp(c_i)} \right)^{W_{its}}
  \quad \times \left( \frac{\exp(c_i)}
  {\exp(E[v_i | I_{it}]) + \exp(c_i)} \right)^{W_{itd} - W_{its}}
\end{align}
where $\bm{\theta}$ represents the parameter vector governing the individual's utility function or learning process: $\bm{\theta} = [c_i, \alpha_{0i}, \alpha_1, \alpha_2, \alpha_3, \alpha_4, \alpha_5, \gamma_{0i}, \gamma_1, \gamma_2, \gamma_3, \gamma_4, \gamma_5, \sigma_n^2]$.

Throughout the model setup, heterogeneity in both the utility function and the signal function is captured using observable variables and is assumed to be independent, conditional on these observables. We now relax this assumption by employing a latent class approach \citep{kamakura1989probabilistic}. Specifically, each individual is assumed to have a probability of belonging to a particular latent class, which introduces class-specific constants related to the utility function and the signal function, respectively. This approach effectively captures and accounts for heterogeneity that is not reflected in the observable variables. Additionally, it allows for unobserved dependencies. For instance, a negative correlation between utility gain and learning speed implies that, all else being equal, individuals who learn more quickly (i.e., have lower variance) also tend to derive relatively higher utility gains from using ChatGPT.

We compare models with varying numbers of latent classes using standard model selection criteria and ultimately select a two-class specification. Without loss of generality, we present the model formulation for the case of two latent classes and report the model selection results in Appendix~\ref{app:model_select}. Specifically, for latent class 1, we assign the probability of individuals belonging to this class as $Pr(Class_i = 1) = \frac{1}{1 + \exp(\lambda)}$, with associated parameters $c_i =c$, $\alpha_{0i} = \alpha_{0}$, and $\gamma_{0i} = \gamma_0$. For latent class 2, the probability is given by $Pr(Class_i = 2) = \frac{\exp(\lambda)}{1 + \exp(\lambda)}$, with associated parameters $c_i =c + \Delta c$, $\alpha_{0i} = \alpha_{0} + \Delta \alpha_{0}$ and $\gamma_{0i} = \gamma_0 + \Delta \gamma_0$.

With the incorporation of the latent class structure, the likelihood function can be rewritten accordingly to account for the probabilistic classification of individuals into different latent classes.
\begin{align}
    l_{it}(W_{itd}, W_{its} \mid \bm{\theta'}) = &\Pr(Class_i = 1) * l_{it}(W_{itd}, W_{its} | \bm{\theta'}, Class_i = 1) + \nonumber \\ 
    &\Pr(Class_i = 2) * l_{it}(W_{itd}, W_{its} | \bm{\theta'}, Class_i = 2)
\end{align}
\normalsize
where $\bm{\theta'}$ includes $[c, \Delta c, \alpha_{0}, \Delta \alpha_{0}, \alpha_1, \alpha_2, \alpha_3, \alpha_4, \alpha_5, \gamma_{0}, \Delta \gamma_0, \gamma_1, \gamma_2, \gamma_3, \gamma_4, \gamma_5, \sigma_n^2, \lambda]$.

\section{Estimation and Identification}
\label{sec:est_identification}

\subsection{Model Estimation}

Based on the likelihood for each (user, period) pair, the total likelihood function can be articulated by considering $I$ users, each observed over $T_i$ periods:
\begin{align}
    L(\bm{\theta'}) = \prod_{i = 0}^{I - 1}\prod_{t = 0}^{T_i - 1} l_{it}(W_{itd}, W_{its} \mid \bm{\theta'})
\end{align}

A common approach to estimate the parameters of interest $\bm{\theta'}$ is to find those values that maximize the empirical likelihood \citep{train2009discrete}. However, this approach is not straightforward for the Bayesian learning model given the available data: we can only observe each individual's attributes such as $College_i$, $AgeHigh_i$, $AgeMid_i$, $Male_i$, $White_i$, $LLMOE_i$, $English_i$, and their usage of ChatGPT ($W_{its}$) and traditional search engines ($W_{itd} - W_{its}$) during each period. We aim to derive the unknown $\bm{\theta'}$, but we cannot directly observe users' perceived representative utility for ChatGPT ($E[v_i | I_{it}]$) at each period or the stochastic error terms ($\varepsilon_{it0}$, $\varepsilon_{it1}$). Similar to the mixed logit model, slight modifications can render a fully closed-form expression of the likelihood unfeasible \citep{train2009discrete}. Although there is a direct mathematical relationship between perceived utility and choice probability under the logit model, deriving the parameters of interest from perceived utility lacks a closed-form solution, as we cannot observe the initial belief distribution $N(v_0, \sigma_0^2)$ or the exact signals ($s_{itk}$ and $n_{itk}$) \citep{erdem1996decision}.

To address this challenge, we adopt established practices in Bayesian learning with unobservable signals and employ a maximum simulated likelihood estimation method \citep{wang2024identification, erdem1996decision}. Specifically, signals are drawn from their respective distributions, governed by the current parameter estimates, for each simulation. These signals are then used in the Bayesian updating process to derive perceived utility and subsequently compute the simulated likelihood. After conducting 100 simulations, the average of the simulated likelihoods is used to approximate the likelihood with all the other unknown components, except for $\bm{\theta'}$, integrated out. This average likelihood then serves as the objective function for optimization, enabling us to estimate the parameters that maximize the simulated likelihood.

\subsection{Identification}
In the Bayesian learning process, users learn the true representative utility from signals and gradually update their mean belief from $v_0$ to $v_i$. Following prior literature \citep{wang2024identification, ching2013learning}, $v_0$ and $v_i$ cannot be simultaneously identified, and therefore $v_0$ is normalized to zero to ensure model identifiability. With $v_0$ held constant, the outside option's representative utility $c_i$ is identified based on users' initial choice probabilities between ChatGPT and traditional search engines. A decrease in $c_i$ corresponds to a relative increase in the net utility from using ChatGPT, thereby increasing the likelihood of ChatGPT adoption in the absence of other information.

Leveraging $c_i$, the representative utility of ChatGPT $v_i$ is identified based on users' steady-state choice probabilities. The utility parameters ($\alpha_{0i}, \alpha_1, \alpha_2, \alpha_3, \alpha_4, \alpha_5, \alpha_6, \alpha_7$) are then derived based on the heterogeneous steady-state choice probabilities among individuals with different backgrounds. For example, an increase in $\alpha_1$ implies a greater impact of education on representative utility, making users who receive college education more likely to choose ChatGPT over traditional search engines.

Furthermore, the variance of the signals ($\sigma_{s_i}^2$ and $\sigma_n^2$) is identified through the variation in users' belief updates after receiving different signals. As with the identification of the mean value, we cannot simultaneously identify the prior variance and the signal variance. Thus, the prior variance $\sigma_0^2$ is fixed at $\exp(4)$ to reflect substantial uncertainty that individuals face when they have no prior information about ChatGPT. The core findings are robust to different choices of prior variance values as shown in Appendix~\ref{app:robustness_variance}. As individuals receive varying numbers of news signals over time, differences in how their beliefs evolve provide information about $\sigma_n^2$. Similarly, the signal parameters ($\gamma_{0i}, \gamma_1, \gamma_2, \gamma_3, \gamma_4, \gamma_5, \gamma_6, \gamma_7$) are identified based on the variation in learning speeds across individuals with different backgrounds.

Finally, since individuals exhibit different dynamics depending on their latent class \citep{zheng2020optimizing}, the parameter $\lambda$ that governs the class probability, along with the associated class-specific parameters ($c$, $\Delta c$, $\alpha_{0}$, $\Delta \alpha_{0}$, $\gamma_0$, and $\Delta \gamma_0$), can be identified. Prior to estimating the model with empirical data, a Monte Carlo simulation is conducted to ensure that all parameters can be accurately identified within the proposed model structure and estimation procedure.

\section{Results}

Following the estimation procedure described earlier, we summarize the results in Table~\ref{tab:results}. First, we document a \textit{utility divide} in a real-world setting, showing that individuals with different social characteristics derive different levels of utility from ChatGPT relative to traditional search engines. Regarding age and gender, we find that younger users and males benefit significantly more from ChatGPT, as reflected in higher estimated utility ($\alpha_2 = -1.168$, $\alpha_3 = -0.886$, and $\alpha_4 = 1.403$). Notably, the utility difference between the young group (18 \textasciitilde{} 34) and the middle-aged group (35 \textasciitilde{} 47) is larger than the difference between the middle-aged group and the older group (47+). To further assess the age pattern, we estimate an alternative specification that includes continuous $Age$ and $Age^2$ and report the results in Appendix~\ref{app:robustness_age_2}. The small but positive coefficient on $Age^2$ suggests that as age increases, relative utility declines at a slightly decreasing rate. These findings are consistent with the theoretical argument that younger and male users tend to develop higher AI literacy through day-to-day use, and therefore derive greater utility from generative AI relative to traditional search engines \citep{yan2024promises, robinson2020digital, scheerder2017determinants}.

Further, our model indicates that individuals with higher levels of LLM occupational exposure derive greater utility from ChatGPT ($\alpha_6 = 1.058$). This pattern is consistent with prior digital divide research showing that individuals with more advanced computer skills are better positioned to use ICT effectively \citep{hargittai2003informed, van2003digital}. In the LLM context, higher $LLMOE$ typically implies that users' job tasks are more closely aligned with LLM functionalities and that they have more opportunities to incorporate ChatGPT into routine workflows, thereby realizing higher utility \citep{eloundou2024gpts}.

\begin{table}[htbp] \footnotesize
\centering
\caption{Parameters Estimation}
\label{tab:results}
\begin{tabular}{l p{11cm} c c }
\toprule
\textbf{Parameter} & \textbf{Description} & \textbf{Value} & \textbf{Std} \\
\midrule
$v_0$ & Prior belief on the mean value of representative utility & 0 & -- (fixed) \\
log($\sigma_0^2$) & Prior belief on the variance of representative utility & 4 & -- (fixed) \\
$\lambda$ & Latent class probability parameter & $-$1.336 & $(0.071)^{***}$ \\
$c$ & Mean representative utility of traditional search engines & 0.627 & $(0.008)^{***}$ \\
$\Delta c$ & Difference in mean representative utility of traditional search engines between two latent classes & 0.401 & $(0.017)^{***}$ \\
$\alpha_0$ & Intercept of ChatGPT's representative utility & 0.402 & $(0.055)^{***}$ \\
$log(\Delta\alpha_0)$ & Difference in ChatGPT's representative utility between two latent classes & 0.537 & $(0.021)^{***}$ \\
$\alpha_1(College)$ & Coefficient of college education on ChatGPT's representative utility & $-$0.539 & $(0.015)^{***}$ \\
$\alpha_2(AgeHigh)$ & Coefficient of high age group on ChatGPT's representative utility & $-$1.168 & $(0.022)^{***}$ \\
$\alpha_3(AgeMid)$ & Coefficient of middle age group on ChatGPT's representative utility & $-$0.886 & $(0.019)^{***}$ \\
$\alpha_4(Male)$ & Coefficient of males on ChatGPT's representative utility & 1.403 & $(0.017)^{***}$ \\
$\alpha_5(White)$ & Coefficient of race (white vs non-white) on ChatGPT's representative utility & $-$0.900 & $(0.011)^{***}$ \\
$\alpha_6(LLMOE)$ & Coefficient of LLM occupational exposure on ChatGPT's representative utility  & 1.058 & $(0.028)^{***}$ \\
$\alpha_7(English)$ & Coefficient of English literacy on ChatGPT's representative utility  & $-$1.107 & $(0.057)^{***}$ \\
$\sigma_n^2$ & News signal variance & 5.777 & $(0.024)^{***}$ \\
$\gamma_0$ & Intercept of usage signal variance & 5.051 & $(0.061)^{***}$ \\
$\Delta\gamma_0$ & Difference in usage signal variance between two latent classes & 3.748 & $(0.022)^{***}$ \\
$\gamma_1(College)$ & Coefficient of college education on usage signal variance & $-$0.272 & $(0.017)^{***}$ \\
$\gamma_2(AgeHigh)$ & Coefficient of high age group on usage signal variance & 0.288 & $(0.021)^{***}$ \\
$\gamma_3(AgeMid)$ & Coefficient of middle age group on usage signal variance & 0.175 & $(0.020)^{***}$ \\
$\gamma_4(Male)$ & Coefficient of males on usage signal va
riance& $-$0.193 & $(0.017)^{***}$ \\
$\gamma_5(White)$ & Coefficient of white people on usage signal variance & $-$0.712 & $(0.018)^{***}$ \\
$\gamma_6(LLMOE)$ & Coefficient of LLM occupational exposure on usage signal variance & $-$0.925 & $(0.079)^{***}$ \\
$\gamma_7(English)$ & Coefficient of English literacy on usage signal variance & $-$0.757 & $(0.046)^{***}$ \\
\bottomrule
\multicolumn{3}{l}{ \textit{Note}: Standard errors in parentheses; $^{***}$ p$<$0.01, $^{**}$ p$<$0.05, $^{*}$ p$<$0.1.} \\
\end{tabular}
\end{table}

While our findings corroborate prior observations for several social characteristics, they also provide evidence that challenges established views about who benefits more in the digital divide literature. Most notably, with respect to education, one of the most prominent predictors of the digital divide \citep{autor1998computing}, we find that users without any college education obtain substantially higher relative utility from ChatGPT ($\alpha_1 = -0.539$). This result appears to contrast with prior research suggesting that education is positively associated with ICT adoption outcomes \citep{scheerder2017determinants}. However, it can be reconciled by noting that users with some college education may have higher overall literacy and thus can leverage both ChatGPT and traditional search engines more effectively. Because ChatGPT provides natural-language answers that can be easier to interpret, it may function as an ``equalizer'' that users with and without college education can both utilize. As a result, users without college education may benefit more in relative terms because their baseline utility from traditional search engines is lower.

Furthermore, we find that white users obtain comparatively lower utility from ChatGPT ($\alpha_5 = -0.900$), which contrasts with prior work that typically finds white users have greater access to and derive more benefits from digital technologies \citep{fairlie2004race}. In addition, users with lower English literacy gain significantly more utility from ChatGPT relative to search engines ($\alpha_7 = -1.107$). These results suggest another channel through which cultural and language backgrounds shape generative AI use: although the underlying LLM is trained disproportionately on English-language corpora and may align more closely with Western cultural contexts \citep{agarwal2025ai, martinez2025llms, qin2025survey}, its natural-language interaction mode can also help users from other backgrounds articulate their needs in detail and obtain information that may be harder to navigate through traditional search engines \citep{xu2023chatgpt}. As a result, non-white users and those with lower English literacy may experience larger relative gains from ChatGPT compared with traditional search engines.

In addition to the divide in actual utility, we observe a significant deviation of \( v_i \) from \( v_0 \), where \( v_0 \) is normalized to zero. For example, consider a female user younger than 35 who is non-white, did not receive any college education, has zero English literacy and no LLM occupational exposure, and belongs to latent class 2. For this individual, the actual representative utility $v_i$ is underestimated by $ (\alpha_0 + \Delta \alpha)  - v_0= 2.113$. Consequently, without an effective learning process, these users may experience substantial welfare losses due to failure in choosing the more suitable tools. Given the pronounced discrepancy between initial beliefs and actual utility, it is crucial for individuals to learn and adjust their perceptions to enable more informed decision-making.

However, individuals' ability to update these beliefs may vary substantially across social characteristics (\textit{learning divide}), and the groups that learn faster may not be the same as those that experience higher per-use utility gains. By definition, a larger signal variance implies slower learning. This divergence is already evident across the latent classes: holding observable social characteristics constant, a user in latent class 2 obtains a larger gain from ChatGPT ($\Delta\alpha_0 - \Delta c = \exp(0.537) - 0.401 = 1.310$) but learns more slowly ($\Delta\gamma_0 = 3.748$) than a comparable user in latent class 1.

Interestingly, while non-white individuals, users with lower English literacy, and users without college education tend to benefit more from ChatGPT on a per-use basis, they learn significantly more slowly, as reflected in the signal-variance parameters ($\gamma_1 = -0.272$, $\gamma_5 = -0.712$, and $\gamma_7 = -0.757$). Thus, absent targeted interventions, these users may adopt ChatGPT more slowly and may be more prone to extreme belief updates, potentially leading to suboptimal usage patterns. These heterogeneities help reconcile the apparent tension between per-use utility gains and adoption patterns documented in the digital divide literature \citep{scheerder2017determinants, fairlie2004race}: although these users obtain higher utility from each interaction, slower learning can impede efficient adoption of ChatGPT. 

In contrast, younger and male users update toward the true benefits of ChatGPT more quickly ($\gamma_2 = 0.288$, $\gamma_3 = 0.175$, and $\gamma_4 = -0.193$). Consistent with the utility estimates, the relationship between learning speed and age is also nonlinear, with learning speed declining at a decreasing rate as age increases. Likewise, we find that individuals with lower occupational exposure to LLMs exhibit faster learning rates ($\gamma_6 = -0.925$). Across these dimensions, the learning divide partly mirrors the utility divide, thereby amplifying disparities across demographic groups.

Finally, it is important to note that our results remain robust when subjected to: (i) restricting the sample to observations prior to the GPT-4 launch; (ii) alternative value of the prior belief variance (hyperparameter); (iii) the inclusion of different news outlets retrieved in October 2024; (iv) re-estimating the results after the exclusion of outliers; and (v) excluding search queries that are not substitutable by ChatGPT. The results can be found in Appendix~\ref{app:robustness_gpt4}, Appendix~\ref{app:robustness_variance}, Appendix~\ref{app:robustness_alternative_news}, 
Appendix~\ref{app:robustness_excluding_outliers}, and Appendix~\ref{app:robustness_substitution} respectively. These robustness checks further reinforce the validity of our findings.

\section{Policy Simulation}
\label{sec:policy_simulation}

Our empirical results have revealed the presence of both learning and utility divides. While the impact of the utility divide is straightforward, the effect of the learning divide on user adoption outcomes and welfare is more complex. To further elucidate the consequences of the learning divide, including potential negative impacts and possible mitigation strategies (RQ3), we conduct numerical simulations complemented by policy evaluations.

First, to illustrate the role of learning in the adoption process, we simulate the belief trajectory of a typical slow learner. Specifically, we consider a 35-to-47-year-old female non-white user with no English literacy, no college education, and no LLM occupational exposure, who belongs to latent class 2. This individual faces the choice between ChatGPT and traditional search engines 16 times per day\footnote{We use the sample mean of daily activities to represent typical behavior.} over a horizon of 1{,}000 days. The simulated belief path is reported in Appendix~\ref{app:counterfactual}. Despite substantial uncertainty about ChatGPT's true utility, the user quickly finds continued use unrewarding and instead relies on search engines. As a result, she forgoes opportunities to update her beliefs through further interactions and to potentially benefit from ChatGPT. More generally, a small number of negative experiences can generate persistently pessimistic beliefs. We refer to this phenomenon as a \emph{belief trap}, in which users become anchored by negative beliefs, discontinue engagement with the new technology, and are unlikely to escape without external intervention.

Given the important implications of the belief trap, we next examine how the learning divide affects the likelihood of becoming trapped. In addition to the slow learner, we consider a representative fast learner for comparison: a male user younger than 35 who is white, has a college education, and has the highest English literacy and LLM occupational exposure. To control for unobserved heterogeneity, we assign this individual to latent class 2 as well. We define ``being trapped'' as follows: after one year, a user is classified as trapped if their current probability of using ChatGPT, based on their prevailing beliefs, falls below $1\%$ of the optimal probability implied by the true mean utility. We then simulate 10{,}000 belief trajectories for each user and compute the probability of becoming trapped; the resulting trap rates are reported in Appendix~\ref{app:counterfactual}. Our results show that the slow learner is substantially more likely to become trapped, plausibly because greater uncertainty slows learning and makes early negative signals more consequential. To rule out the possibility that this gap is driven by heterogeneous utilities, we conduct an additional counterfactual exercise. Because higher utility increases adoption and thus creates more opportunities for belief updating, we set the slow learner's true utility equal to that of the fast learner and re-run the simulations in a counterfactual scenario. Even under this utility-equalized scenario, slow learners remain more likely to become trapped. This persistent gap in trapping probabilities highlights the role of the learning divide in shaping dynamic adoption behavior and, ultimately, welfare.

Given the substantial impact of the learning divide, mitigating it and its associated consequences becomes crucial. Our simulation results suggest that providing free training may be an effective intervention. In a single illustrative trajectory shown in Figure~\ref{fig:scatter}, we provide training at the outset by ensuring that the slow learner uses ChatGPT 300 times prior to Day~0. Under this intervention, the user is no longer trapped: the belief is updated more frequently and converge toward the true representative utility of ChatGPT. This example illustrates how early training can help users avoid the belief trap; accordingly, offering training to slow learners may mitigate the learning divide. Our large-scale simulations (Figure \ref{fig:error_bar}) further show that a training program comprising roughly 200--400 uses of ChatGPT allows slow learners to achieve trapping probabilities comparable to those of fast learners, suggesting that this approach can be effective in practice. Finally, Appendix~\ref{app:counterfactual} further examines (1) the effects of reading news articles and (2) the effects of potential improvements from more advanced large language models. Overall, the amount of news exposure required to alleviate the generative AI divide appears to be very large relative to the average news consumption in our sample. Likewise, the performance improvements needed for generative AI tools to materially reduce the divide are also substantial. These comparisons suggest that training programs are likely to be the more feasible and cost-effective approach to mitigating the generative AI divide.

\section{Discussion and Conclusion}

\subsection{Summary and Discussion}

Leveraging a large scale clickstream dataset, we develop and estimate a Bayesian learning model to examine generative AI divide and to identify the underlying mechanisms. Several key findings align with those from prior research on digital divide. Consistent with existing evidence, male users, younger users, and individuals in occupations with greater exposure to LLM learn about generative AI's utility more rapidly and obtain higher utility per interaction \citep{mcelheran2024ai, robinson2020digital, scheerder2017determinants, dewan2005digital, van2003digital}. Establishing that these patterns persist in the generative AI context is theoretically important, as it shows that familiar forms of inequality remain even when interaction is mediated by advanced AI systems.

More importantly, our results reveal that, conditional on use, users with some college education, users from marginalized racial groups (i.e., non-white users), and users with higher English literacy obtain lower per-use utility. One possible explanation is that generative AI substitutes for users' baseline cognitive, linguistic, and informational skills rather than complementing them \citep{noy2023experimental, haslberger2023no, autor1998computing}. Because users with some college education and stronger linguistic or cultural fit possess higher baseline capabilities, they can extract utility in both search engine and generative AI settings, leaving less space for incremental gains from synthesized AI outputs \citep{agrawal2024artificial, humlum2024adoption}. By contrast, groups that appear disadvantaged along traditional dimensions may realize higher per-use utility because AI-generated synthesized responses substantially reduce search, filtering, and integration costs relative to the search engine environment \citep{burtch2019investigating, scheerder2017determinants, hargittai2002second}. This interpretation is consistent with emerging evidence that generative AI can disproportionately benefit lower-skilled or less experienced users by substituting for baseline capabilities, while delivering smaller marginal gains for higher-skilled users who have stronger outside options \citep{humlum2024adoption, yan2024promises, noy2023experimental}. This asymmetric value creation generates a divergence between learning speed and realized utility per-use: some socially advantaged users learn system capabilities more quickly yet obtain smaller benefits per interaction \citep{vassilakopoulou2023bridging, lythreatis2022digital}.

\subsection{Contributions and Implications}

This study contributes to the literature on the digital divide, AI adoption, and Bayesian learning in several ways. First, we develop a new framework for understanding AI adoption in environments characterized by informational uncertainty and dynamic learning. Whereas prior work on the digital divide often emphasizes static access and initial use, our approach highlights the ongoing learning dynamics that shape how users interact with AI technologies over time. Leveraging real-world behavioral data and a structural Bayesian learning model, we identify self-reinforcing cycles such as the belief trap and disentangle heterogeneity in belief updating from heterogeneity in per-use utility, a distinction that is difficult to recover from survey- or experiment-based designs. Second, we advance the literature on the digital divide by introducing the concepts of the \emph{learning divide}, \emph{utility divide}, and \emph{belief trap} to clarify the mechanisms behind differential AI adoption and outcomes. In doing so, we extend the digital divide frameworks to incorporate dynamic, path-dependent inequalities arising from learning and belief updating under uncertainty. Third, we show that AI technologies, absent appropriate interventions, may amplify existing social inequalities, underscoring the importance of inclusive design and deployment practices.

Beyond these theoretical contributions, our findings also have practical implications for governments, Non-profit organizations (NGOs), AI model providers, and firms. Governments and NGOs can strengthen users’ capabilities to use AI through educational programs that help underprivileged groups overcome information asymmetry. There is an opportunity to go beyond basic access by offering dynamic ``on-ramping'' supports that reduce signal noise for slow learners. They can consider providing disadvantaged groups with subsidized access to premium, high-fidelity models to reduce the frequency of hallucinations, alongside localized benchmarking tools that help non-native speakers evaluate unstructured outputs more effectively. AI model platforms could also support adoption by facilitating third party evaluations that assess performance across diverse, general-purpose tasks, making system capabilities easier to learn and compare. For organizations providing AI applications to their employees, they have the option to formalize ``AI Navigator'' peer networks where experienced users share prompt-to-output pairs, transforming private learning into social signals that lower uncertainty for others. Firms that provide AI or embed AI into their products can prioritize inclusive design and offer training programs targeted at disadvantaged users. By addressing the utility divide, learning divide, and belief trap, these stakeholders can promote AI adoption and help ensure that the benefits of AI technologies are broadly shared across society.

\subsection{Limitations and Future Work}

Our work also highlights several promising directions for future research. First, as in many studies based on clickstream data or online surveys, the social characteristics in our dataset are self-reported and measured with limited granularity (e.g., a binary education indicator). Although the company seeks to balance attributes during panel recruitment, the resulting panel may not perfectly match the U.S.\ population. In addition, the company cannot guarantee that panelists do not use untracked devices. Since we restrict the sample to individuals with sufficiently high activity levels (at least 60 active days during the observation window), this concern may be partially mitigated, but not eliminated. Despite these data constraints, we hope our study provides a ``half answer'' to a ``big question'' and motivates further work on generative AI adoption and digital divide issues \citep{hosanagar2017senior}. Future research could apply our framework in other generative AI adoption contexts (e.g., within enterprises) to assess the generalizability of our findings and identify important patterns that vary across scenarios.

Second, generative AI products can undergo updates, and existing Bayesian learning approaches typically do not directly accommodate such non-stationarity \citep{wang2024identification, erdem1996decision}. In this study, given the currently available toolkit, we conduct robustness checks that focus on the period before the launch of GPT-4 to reduce concerns about version changes (Appendix \ref{app:robustness_gpt4}). Nonetheless, it would be highly valuable for future methodological work to develop models that explicitly incorporate user learning in the presence of product upgrades, and for future empirical research to leverage such tools to study adoption dynamics more directly.

Furthermore, our Bayesian learning model could be applied to other emerging technologies characterized by uncertainty and learning dynamics, such as blockchain \citep{gao2021blockchain}, to determine whether similar divides exist. Finally, future research could explore how disparities in AI adoption affect market competition and economic inequalities. We call for more interdisciplinary research to further understand these issues and to develop strategies that promote equitable AI adoption, ultimately informing policymakers and organizations in fostering inclusive technological progress.

\bibliography{thebibliography}

@article{burton2023misq,
  title={MISQ's DEI initiatives: A continuing journey.},
  author={Burton-Jones, Andrew and Sarker, Saonee},
  journal={Learned Publishing},
  volume={36},
  number={1},
  year={2023}
}

@article{simchi2020editor,
  title={From the editor: Diversity, equity, and inclusion in management science},
  author={Simchi-Levi, David},
  journal={Management Science},
  volume={66},
  number={9},
  pages={3802--3802},
  year={2020},
  publisher={INFORMS}
}

@article{venkatesh2003user,
  title={User acceptance of information technology: Toward a unified view},
  author={Venkatesh, Viswanath and Morris, Michael G and Davis, Gordon B and Davis, Fred D},
  journal={MIS Quarterly},
  pages={425--478},
  year={2003},
  publisher={JSTOR}
}

@article{wang2024identification,
  title={Identification of Structural Learning Models},
  author={Wang, Zhide and Yang, Nathan},
  journal={Available at SSRN 4906492},
  year={2024}
}

@article{liu2024datasets,
  title={Datasets for Large Language Models: A Comprehensive Survey},
  author={Liu, Yang and Cao, Jiahuan and Liu, Chongyu and Ding, Kai and Jin, Lianwen},
  journal={arXiv preprint arXiv:2402.18041},
  year={2024}
}

@article{kojima2022large,
  title={Large language models are zero-shot reasoners},
  author={Kojima, Takeshi and Gu, Shixiang Shane and Reid, Machel and Matsuo, Yutaka and Iwasawa, Yusuke},
  journal={Advances in neural information processing systems},
  volume={35},
  pages={22199--22213},
  year={2022}
}

@article{morris2023levels,
  title={Levels of AGI: Operationalizing Progress on the Path to AGI},
  author={Morris, Meredith Ringel and Sohl-dickstein, Jascha and Fiedel, Noah and Warkentin, Tris and Dafoe, Allan and Faust, Aleksandra and Farabet, Clement and Legg, Shane},
  journal={arXiv preprint arXiv:2311.02462},
  year={2023}
}

@article{xu2023chatgpt,
  title={ChatGPT vs. Google: a comparative study of search performance and user experience},
  author={Xu, Ruiyun and Feng, Yue and Chen, Hailiang},
  journal={arXiv preprint arXiv:2307.01135},
  year={2023}
}

@article{ouyang2022training,
  title={Training language models to follow instructions with human feedback},
  author={Ouyang, Long and Wu, Jeffrey and Jiang, Xu and Almeida, Diogo and Wainwright, Carroll and Mishkin, Pamela and Zhang, Chong and Agarwal, Sandhini and Slama, Katarina and Ray, Alex and others},
  journal={Advances in neural information processing systems},
  volume={35},
  pages={27730--27744},
  year={2022}
}

@article{liu2025deepseek,
  title={Deepseek-v3. 2: Pushing the frontier of open large language models},
  author={Liu, Aixin and Mei, Aoxue and Lin, Bangcai and Xue, Bing and Wang, Bingxuan and Xu, Bingzheng and Wu, Bochao and Zhang, Bowei and Lin, Chaofan and Dong, Chen and others},
  journal={arXiv preprint arXiv:2512.02556},
  year={2025}
}

@inproceedings{zhong2024agieval,
  title={Agieval: A human-centric benchmark for evaluating foundation models},
  author={Zhong, Wanjun and Cui, Ruixiang and Guo, Yiduo and Liang, Yaobo and Lu, Shuai and Wang, Yanlin and Saied, Amin and Chen, Weizhu and Duan, Nan},
  booktitle={Findings of the Association for Computational Linguistics: NAACL 2024},
  pages={2299--2314},
  year={2024}
}

@article{zhao20247b,
  title={7B Fully Open Source Moxin-LLM--From Pretraining to GRPO-based Reinforcement Learning Enhancement},
  author={Zhao, Pu and Shen, Xuan and Kong, Zhenglun and Shen, Yixin and Chang, Sung-En and Rupprecht, Timothy and Lu, Lei and Nan, Enfu and Yang, Changdi and He, Yumei and others},
  journal={arXiv preprint arXiv:2412.06845},
  year={2024}
}

@article{noy2023experimental,
  title={Experimental evidence on the productivity effects of generative artificial intelligence},
  author={Noy, Shakked and Zhang, Whitney},
  journal={Science},
  volume={381},
  number={6654},
  pages={187--192},
  year={2023},
  publisher={American Association for the Advancement of Science}
}

@article{ng2021conceptualizing,
  title={Conceptualizing AI literacy: An exploratory review},
  author={Ng, Davy Tsz Kit and Leung, Jac Ka Lok and Chu, Samuel Kai Wah and Qiao, Maggie Shen},
  journal={Computers and Education: Artificial Intelligence},
  volume={2},
  pages={100041},
  year={2021},
  publisher={Elsevier}
}

@article{mcelheran2024ai,
  title={AI adoption in America: Who, what, and where},
  author={McElheran, Kristina and Li, J Frank and Brynjolfsson, Erik and Kroff, Zachary and Dinlersoz, Emin and Foster, Lucia and Zolas, Nikolas},
  journal={Journal of Economics \& Management Strategy},
  year={2024},
  publisher={Wiley Online Library}
}

@article{burtch2019investigating,
  title={Investigating the relationship between medical crowdfunding and personal bankruptcy in the United States: Evidence of a digital divide},
  author={Burtch, Gordon and Chan, Jason},
  journal={MIS Quarterly},
  volume={43},
  number={1},
  pages={237--262},
  year={2019}
}

@article{choudhury2023investigating,
  title={Investigating the impact of user trust on the adoption and use of ChatGPT: Survey analysis},
  author={Choudhury, Amrit and Shamszare, Hamed},
  journal={Journal of Medical Internet Research},
  volume={25},
  pages={e47184},
  year={2023}
}

@article{gupta2024adoption,
  title={Adoption and impacts of generative artificial intelligence: Theoretical underpinnings and research agenda},
  author={Gupta, Ramesh and Nair, Kavita and Mishra, Mukesh and Ibrahim, Bola and Bhardwaj, Sunita},
  journal={International Journal of Information Management Data Insights},
  volume={4},
  number={1},
  pages={100232},
  year={2024}
}

@article{kohli2024digital,
  title={The digital divide in access to broadband internet and mental healthcare},
  author={Kohli, Kushal and Jain, Bhavya and Patel, Tanmay A. and Eken, Halil N. and Dee, Edmund C. and Torous, John},
  journal={Nature Mental Health},
  pages={1--8},
  year={2024}
}

@article{ma2024exploring,
  title={Exploring User Adoption of ChatGPT: A Technology Acceptance Model Perspective},
  author={Ma, Jianqing and Wang, Ping and Li, Bin and Wang, Tao and Pang, Xiao Sheng and Wang, Dan},
  journal={International Journal of Human–Computer Interaction},
  pages={1--15},
  year={2024}
}

@techreport{oecd2001understanding,
  title={Understanding the Digital Divide},
  author={OECD},
  institution={OECD Publishing},
  number={49},
  year={2001},
  type={OECD Digital Economy Papers}
}

@article{vassilakopoulou2023bridging,
  title={Bridging digital divides: A literature review and research agenda for information systems research},
  author={Vassilakopoulou, Polyxeni and Hustad, Eli},
  journal={Information Systems Frontiers},
  volume={25},
  number={3},
  pages={955--969},
  year={2023}
}

@article{wei2011conceptualizing,
  title={Conceptualizing and testing a social cognitive model of the digital divide},
  author={Wei, Kar Yan and Teo, Hock Hai and Chan, Hock Chuan and Tan, Bernard C. Y.},
  journal={Information Systems Research},
  volume={22},
  number={1},
  pages={170-187},
year={2011}
}

@article{yang2023beyond,
title={Beyond structural inequality: a socio-technical approach to the digital divide in the platform environment},
author={Yang, Jun and Zhang, Min},
journal={Humanities and Social Sciences Communications},
volume={10},
number={1},
pages={1--12},
year={2023}
}

@article{wang2024artificial,
  title={The artificial intelligence divide: Who is the most vulnerable?},
  author={Wang, Chen and Boerman, Sophie C. and Kroon, Anne C. and Möller, Judith and de Vreese, Claes H.},
  journal={New Media \& Society},
  pages={14614448241232345},
  year={2024}
}

@article{jo2023analyzing,
  title={Analyzing ChatGPT adoption drivers with the TOEK framework},
  author={Jo, Hyunjin and Bang, Yongseok},
  journal={Scientific Reports},
  volume={13},
  number={1},
  pages={22606},
  year={2023},
  publisher={Nature Publishing Group}
}

@article{prasad2023towards,
  title={Towards adoption of generative AI in organizational settings},
  author={Prasad Agrawal, K.},
  journal={Journal of Computer Information Systems},
  pages={1--16},
  year={2023}
}

@article{xu2024hallucination,
  title={Hallucination is inevitable: An innate limitation of large language models},
  author={Xu, Ziwei and Jain, Sanjay and Kankanhalli, Mohan},
  journal={arXiv preprint arXiv:2401.11817},
  year={2024}
}

@article{liu2023generate,
  title={" Generate" the Future of Work through AI: Empirical Evidence from Online Labor Markets},
  author={Liu, Jin and Xu, Xingchen and Li, Yongjun and Tan, Yong},
  journal={arXiv preprint arXiv:2308.05201},
  year={2023}
}

@article{humlum2024adoption,
  title={The Adoption of ChatGPT},
  author={Humlum, Anders and Vestergaard, Emilie},
  journal={University of Chicago, Becker Friedman Institute for Economics Working Paper},
  number={2024-50},
  year={2024}
}

@article{huang2014crowdsourcing,
  title={Crowdsourcing new product ideas under consumer learning},
  author={Huang, Yan and Vir Singh, Param and Srinivasan, Kannan},
  journal={Management science},
  volume={60},
  number={9},
  pages={2138--2159},
  year={2014},
  publisher={INFORMS}
}

@article{zhang2020learning,
  title={Learning individual behavior using sensor data: The case of global positioning system traces and taxi drivers},
  author={Zhang, Yingjie and Li, Beibei and Krishnan, Ramayya},
  journal={Information Systems Research},
  volume={31},
  number={4},
  pages={1301--1321},
  year={2020},
  publisher={INFORMS}
}

@article{erdem1996decision,
  title={Decision-making under uncertainty: Capturing dynamic brand choice processes in turbulent consumer goods markets},
  author={Erdem, T{\"u}lin and Keane, Michael P},
  journal={Marketing Science},
  volume={15},
  number={1},
  pages={1--20},
  year={1996},
  publisher={INFORMS}
}

@article{ching2013learning,
  title={Learning models: An assessment of progress, challenges, and new developments},
  author={Ching, Andrew T and Erdem, T{\"u}lin and Keane, Michael P},
  journal={Marketing Science},
  volume={32},
  number={6},
  pages={913--938},
  year={2013},
  publisher={INFORMS}
}

@book{train2009discrete,
  title={Discrete choice methods with simulation},
  author={Train, Kenneth E},
  year={2009},
  publisher={Cambridge university press}
}

@article{fang2022effects,
  title={The effects of online review platforms on restaurant revenue, consumer learning, and welfare},
  author={Fang, Limin},
  journal={Management Science},
  volume={68},
  number={11},
  pages={8116--8143},
  year={2022},
  publisher={INFORMS}
}

@article{lu2022microblogging,
  title={MICROBLOGGING REPLIES AND OPINION POLARIZATION: A NATURAL EXPERIMENT.},
  author={Lu, Yingda and Wu, Junjie and Tan, Yong and Chen, Jian},
  journal={MIS Quarterly},
  volume={46},
  number={4},
  year={2022}
}

@article{van2021bayesian,
  title={Bayesian statistics and modelling},
  author={van de Schoot, Rens and Depaoli, Sarah and King, Ruth and Kramer, Bianca and M{\"a}rtens, Kaspar and Tadesse, Mahlet G and Vannucci, Marina and Gelman, Andrew and Veen, Duco and Willemsen, Joukje and others},
  journal={Nature Reviews Methods Primers},
  volume={1},
  number={1},
  pages={1},
  year={2021},
  publisher={Nature Publishing Group UK London}
}

@book{degroot2005optimal,
  title={Optimal statistical decisions},
  author={DeGroot, Morris H},
  year={2005},
  publisher={John Wiley \& Sons}
}

@article{zhou2022learning,
  title={Learning to be Proficient? A Structural Model of User Dynamic Engagement in E-Health Interventions},
  author={Zhou, Tongxin and Wang, Yingfei and Yan, Lu Lucy and Tan, Yong},
  journal={Available at SSRN 4066017},
  year={2022}
}

@article{haslberger2023no,
  title={No great equalizer: experimental evidence on AI in the UK labor market},
  author={Haslberger, M. and Gingrich, J. and Bhatia, J.},
  journal={Available at SSRN 4594466},
  year={2023}
}

@book{banerjee2011poor,
  title={Poor Economics: A Radical Rethinking of the Way to Fight Global Poverty},
  author={Banerjee, Abhijit},
  year={2011},
  publisher={PublicAffairs}
}

@article{celik2023exploring,
  title={Exploring the determinants of artificial intelligence (AI) literacy: Digital divide, computational thinking, cognitive absorption},
  author={Celik, I.},
  journal={Telematics and Informatics},
  volume={83},
  pages={102026},
  year={2023},
  publisher={Elsevier}
}

@article{erdem2008dynamic,
  title={A dynamic model of brand choice when price and advertising signal product quality},
  author={Erdem, T{\"u}lin and Keane, Michael P and Sun, Baohong},
  journal={Marketing Science},
  volume={27},
  number={6},
  pages={1111--1125},
  year={2008},
  publisher={INFORMS}
}

@article{lin2015learning,
  title={Learning from experience, simply},
  author={Lin, Song and Zhang, Juanjuan and Hauser, John R},
  journal={Marketing Science},
  volume={34},
  number={1},
  pages={1--19},
  year={2015},
  publisher={INFORMS}
}

@article{ching2020structural,
  title={A structural model of correlated learning and late-mover advantages: The case of statins},
  author={Ching, Andrew T and Lim, Hyunwoo},
  journal={Management Science},
  volume={66},
  number={3},
  pages={1095--1123},
  year={2020},
  publisher={INFORMS}
}

@article{chen2020dynamic,
  title={A dynamic model of rational addiction with stockpiling and learning: An empirical examination of e-cigarettes},
  author={Chen, Jialie and Rao, Vithala R},
  journal={Management Science},
  volume={66},
  number={12},
  pages={5886--5905},
  year={2020},
  publisher={INFORMS}
}

@article{zhou2021disclosure,
  title={Disclosure dynamics and investor learning},
  author={Zhou, Frank S},
  journal={Management Science},
  volume={67},
  number={6},
  pages={3429--3446},
  year={2021},
  publisher={INFORMS}
}

@article{zheng2020optimizing,
  title={Optimizing two-sided promotion for transportation network companies: A structural model with conditional Bayesian learning},
  author={Zheng, Jinyang and Ren, Fei and Tan, Yong and Chen, Xi},
  journal={Information Systems Research},
  volume={31},
  number={3},
  pages={692--714},
  year={2020},
  publisher={INFORMS}
}

@article{ho2017disconfirmation,
  title={Disconfirmation effect on online rating behavior: A structural model},
  author={Ho, Yi-Chun and Wu, Junjie and Tan, Yong},
  journal={Information Systems Research},
  volume={28},
  number={3},
  pages={626--642},
  year={2017},
  publisher={INFORMS}
}

@article{ghose2011empirical,
  title={An empirical analysis of user content generation and usage behavior on the mobile Internet},
  author={Ghose, Anindya and Han, Sang Pil},
  journal={Management Science},
  volume={57},
  number={9},
  pages={1671--1691},
  year={2011},
  publisher={INFORMS}
}

@article{scheerder2017determinants,
  title={Determinants of Internet skills, uses and outcomes: A systematic review of the second- and third-level digital divide},
  author={Scheerder, A. and van Deursen, A. and van Dijk, J.},
  journal={Telematics and Informatics},
  volume={34},
  number={8},
  pages={1607--1624},
  year={2017},
  publisher={Elsevier}
}

@article{lythreatis2022digital,
  title={The digital divide: A review and future research agenda},
  author={Lythreatis, S. and Singh, S. K. and El-Kassar, A. N.},
  journal={Technological Forecasting and Social Change},
  volume={175},
  pages={121359},
  year={2022},
  publisher={Elsevier}
}

@article{dewan2005digital,
  title={The digital divide: Current and future research directions},
  author={Dewan, S. and Riggins, F. J.},
  journal={Journal of the Association for Information Systems},
  volume={6},
  number={12},
  pages={298--337},
  year={2005},
  publisher={Association for Information Systems}
}

@article{jackson2001racial,
  title={The Racial Digital Divide: Motivational, Affective, and Cognitive Correlates of Internet Use},
  author={Jackson, L. A. and Ervin, K. S. and Gardner, P. D. and Schmitt, N.},
  journal={Journal of Applied Social Psychology},
  volume={31},
  number={10},
  pages={2019--2046},
  year={2001},
  publisher={Wiley}
}

@article{hargittai2002second,
  title={Second-order Digital Divide: Differences in People's Online Skills},
  author={Hargittai, E.},
  journal={First Monday},
  volume={7},
  number={4},
  year={2002}
}

@article{van2003digital,
  title={The digital divide as a complex and dynamic phenomenon},
  author={Van Dijk, J. and Hacker, K.},
  journal={The Information Society},
  volume={19},
  number={4},
  pages={315--326},
  year={2003},
  publisher={Taylor \& Francis}
}

@incollection{hargittai2003informed,
  title={Informed Web Surfing: The Social Context of User Sophistication},
  author={Hargittai, E.},
  booktitle={Society Online: The Internet in Context},
  year={2003},
  publisher={SAGE Publications}
}

@techreport{fairlie2004race,
  title={Race and the Digital Divide},
  author={Fairlie, R. W.},
  year={2004},
  institution={University of California, Santa Cruz}
}

@article{hsieh2008understanding,
  title={Understanding Digital Inequality: Comparing Continued Use Behavioral Models of the Socio-economically Advantaged and Disadvantaged},
  author={Hsieh, J. J. P.-A. and Rai, A. and Keil, M.},
  journal={MIS Quarterly},
  volume={32},
  number={1},
  pages={97--126},
  year={2008},
  publisher={Management Information Systems Research Center, University of Minnesota}
}

@article{venkatesh2014understanding,
  title={Understanding e‐Government portal use in rural India: role of demographic and personality characteristics},
  author={Venkatesh, V. and Sykes, T. A. and Venkatraman, S.},
  journal={Information Systems Journal},
  volume={24},
  number={3},
  pages={249--269},
  year={2014},
  publisher={Wiley}
}

@article{boeing2020online,
  title={Online rental housing market representation and the digital reproduction of urban inequality},
  author={Boeing, G.},
  journal={Environment and Planning A: Economy and Space},
  volume={52},
  number={2},
  pages={449--468},
  year={2020},
  publisher={SAGE Publications}
}

@article{robinson2020digital,
  title={Digital inequalities 2.0: Legacy inequalities in the information age},
  author={Robinson, L. and others},
  journal={First Monday},
  volume={25},
  number={7},
  year={2020}
}

@article{hidalgo2020digital,
  title={The digital divide in light of sustainable development: An approach through advanced machine learning techniques},
  author={Hidalgo, A. and Gabaly, S. and Morales-Alonso, G. and Urueña, A.},
  journal={Technological Forecasting and Social Change},
  volume={150},
  pages={119754},
  year={2020},
  publisher={Elsevier}
}

@article{hsieh2011addressing,
  title={Addressing Digital Inequality for the Socioeconomically Disadvantaged Through Government Initiatives: Forms of Capital That Affect ICT Utilization},
  author={Hsieh, J. J. P.-A. and Rai, A. and Keil, M.},
  journal={Information Systems Research},
  volume={22},
  number={2},
  pages={233--253},
  year={2011},
  publisher={INFORMS}
}

@article{autor1998computing,
  title={Computing Inequality: Have Computers Changed the Labor Market?},
  author={Autor, David H. and Katz, Lawrence F. and Krueger, Alan B.},
  journal={The Quarterly Journal of Economics},
  volume={113},
  number={4},
  pages={1169--1213},
  year={1998},
  publisher={Oxford University Press}
}

@article{agrawal2024artificial,
  title={Artificial intelligence adoption and system‐wide change},
  author={Agrawal, A. and Gans, J. S. and Goldfarb, A.},
  journal={Journal of Economics \& Management Strategy},
  volume={33},
  number={2},
  pages={327--337},
  year={2024},
  publisher={Wiley Online Library}
}

@article{alekseeva2020ai,
  title={AI Adoption and firm performance: Management versus IT},
  author={Alekseeva, L. and Gine, M. and Samila, S. and Taska, B.},
  journal={Available at SSRN 3677237},
  year={2020}
}

@article{gans2023artificial,
  title={Artificial intelligence adoption in a monopoly market},
  author={Gans, J. S.},
  journal={Managerial and Decision Economics},
  volume={44},
  number={2},
  pages={1098--1106},
  year={2023}
}

@article{chen2024does,
  title={How does worker mobility affect business adoption of a new technology? The case of machine learning},
  author={Chen, R. and Balasubramanian, N. and Forman, C.},
  journal={Strategic Management Journal},
  year={2024}
}

@article{kamakura1989probabilistic,
  title={A probabilistic choice model for market segmentation and elasticity structure},
  author={Kamakura, Wagner A and Russell, Gary J},
  journal={Journal of marketing research},
  volume={26},
  number={4},
  pages={379--390},
  year={1989},
  publisher={SAGE Publications Sage CA: Los Angeles, CA}
}

@article{wang2023human,
  title={Human-AI co-creation in product ideation: The dual view of quality and diversity},
  author={Wang, Wen and Yang, Mochen and Sun, Tianshu},
  journal={Available at SSRN 4668241},
  year={2023}
}

@article{gao2021blockchain,
  title={Blockchain Technology Adoption in Digital Advertising: A Game-Theoretic Model},
  author={Gao, Yi and Kumar, Subodha and Liu, Dengpan},
  journal={Available at SSRN 3640784},
  year={2021}
}

@article{hu2022human,
  title={Human-algorithmic bias: Source, evolution, and impact},
  author={Hu, Xiyang and Huang, Yan and Li, Beibei and Lu, Tian},
  journal={Available at SSRN 4195014},
  year={2022}
}

@article{zhang2024regurgitative,
  title={Regurgitative training: The value of real data in training large language models},
  author={Zhang, Jinghui and Qiao, Dandan and Yang, Mochen and Wei, Qiang},
  journal={arXiv preprint arXiv:2407.12835},
  year={2024}
}

@article{yan2024promises,
  title={Promises and challenges of generative artificial intelligence for human learning},
  author={Yan, Lixiang and Greiff, Samuel and Teuber, Ziwen and Ga{\v{s}}evi{\'c}, Dragan},
  journal={Nature Human Behaviour},
  volume={8},
  number={10},
  pages={1839--1850},
  year={2024},
  publisher={Nature Publishing Group UK London}
}

@article{annapureddy2025generative,
  title={Generative AI literacy: Twelve defining competencies},
  author={Annapureddy, Ravinithesh and Fornaroli, Alessandro and Gatica-Perez, Daniel},
  journal={Digital Government: Research and Practice},
  volume={6},
  number={1},
  pages={1--21},
  year={2025},
  publisher={ACM New York, NY}
}

@article{wang2025artificial,
  title={Artificial Intelligence (AI) Assistant in Online Shopping: A Randomized Field Experiment on a Livestream Selling Platform},
  author={Wang, Lingli and Huang, Ni and He, Yumei and Liu, De and Guo, Xunhua and Sun, Yan and Chen, Guoqing},
  journal={Information Systems Research},
  year={2025},
  publisher={INFORMS}
}

@article{eloundou2024gpts,
  title={GPTs are GPTs: Labor market impact potential of LLMs},
  author={Eloundou, Tyna and Manning, Sam and Mishkin, Pamela and Rock, Daniel},
  journal={Science},
  volume={384},
  number={6702},
  pages={1306--1308},
  year={2024},
  publisher={American Association for the Advancement of Science}
}

@article{zheng2012business,
  title={From business intelligence to competitive intelligence: Inferring competitive measures using augmented site-centric data},
  author={Zheng, Zhiqiang and Fader, Peter and Padmanabhan, Balaji},
  journal={Information Systems Research},
  volume={23},
  number={3-part-1},
  pages={698--720},
  year={2012},
  publisher={INFORMS}
}

@article{chiou2022restrictions,
  title={How do restrictions on advertising affect consumer search?},
  author={Chiou, Lesley and E. Tucker, Catherine},
  journal={Management Science},
  volume={68},
  number={2},
  pages={866--882},
  year={2022},
  publisher={INFORMS}
}

@article{chesnes2017banning,
  title={Banning foreign pharmacies from sponsored search: The online consumer response},
  author={Chesnes, Matthew and Dai, Weijia and Zhe Jin, Ginger},
  journal={Marketing Science},
  volume={36},
  number={6},
  pages={879--907},
  year={2017},
  publisher={INFORMS}
}

@inproceedings{bassignana2025ai,
  title={The AI gap: How socioeconomic status affects language technology interactions},
  author={Bassignana, Elisa and Curry, Amanda Cercas and Hovy, Dirk},
  booktitle={Proceedings of the 63rd Annual Meeting of the Association for Computational Linguistics (Volume 1: Long Papers)},
  pages={18647--18664},
  year={2025}
}

@article{qin2025survey,
  title={A survey of multilingual large language models},
  author={Qin, Libo and Chen, Qiguang and Zhou, Yuhang and Chen, Zhi and Li, Yinghui and Liao, Lizi and Li, Min and Che, Wanxiang and Yu, Philip S},
  journal={Patterns},
  volume={6},
  number={1},
  year={2025},
  publisher={Elsevier}
}

@article{martinez2025llms,
  title={Do LLMs exhibit the same commonsense capabilities across languages?},
  author={Mart{\'\i}nez-Murillo, Ivan and Lloret, Elena and Moreda, Paloma and Gatt, Albert},
  journal={arXiv preprint arXiv:2509.06401},
  year={2025}
}

@inproceedings{broder2002taxonomy,
  title={A taxonomy of web search},
  author={Broder, Andrei},
  booktitle={ACM Sigir forum},
  volume={36},
  number={2},
  pages={3--10},
  year={2002},
  organization={ACM New York, NY, USA}
}

@article{sun2024trusting,
  title={Trusting the search: unraveling human trust in health information from Google and ChatGPT},
  author={Sun, Xin and Ma, Rongjun and Zhao, Xiaochang and Li, Zhuying and Lindqvist, Janne and Ali, Abdallah El and Bosch, Jos A},
  journal={arXiv preprint arXiv:2403.09987},
  year={2024}
}

@inproceedings{karunaratne2023new,
  title={Is it the new Google: Impact of ChatGPT on students’ information search habits},
  author={Karunaratne, Thashmee and Adesina, Adenike},
  booktitle={Proceedings of the 22nd European Conference on e-Learning, ECEL},
  pages={147--155},
  year={2023}
}

@article{salthouse2012consequences,
  title={Consequences of age-related cognitive declines},
  author={Salthouse, Timothy},
  journal={Annual Review of Psychology},
  volume={63},
  number={1},
  pages={201--226},
  year={2012},
  publisher={Annual Reviews}
}

@inproceedings{agarwal2025ai,
  title={AI suggestions homogenize writing toward western styles and diminish cultural nuances},
  author={Agarwal, Dhruv and Naaman, Mor and Vashistha, Aditya},
  booktitle={Proceedings of the 2025 CHI Conference on Human Factors in Computing Systems},
  pages={1--21},
  year={2025}
}

@article{huang2019level,
  title={“Level up”: Leveraging skill and engagement to maximize player game-play in online video games},
  author={Huang, Yan and Jasin, Stefanus and Manchanda, Puneet},
  journal={Information Systems Research},
  volume={30},
  number={3},
  pages={927--947},
  year={2019},
  publisher={INFORMS}
}

@misc{hosanagar2017senior,
  title={Senior editor perspectives},
  author={Hosanagar, Kartik},
  journal={Information Systems Research},
  volume={28},
  number={4},
  pages={689--689},
  year={2017},
  publisher={INFORMS}
}
\bibliographystyle{informs2014}

\newpage
\begin{APPENDICES}

\renewcommand{\thesection}{\Alph{section}}
\renewcommand{\thesubsection}{\Alph{section}.\arabic{subsection}}
\counterwithin{table}{section}
\counterwithin{figure}{section}
\renewcommand{\thetable}{\Alph{section}\arabic{table}}
\renewcommand{\thefigure}{\Alph{section}\arabic{figure}}

\begin{landscape}
\section{Literature}

\scriptsize
\begin{longtable}{
  >{\raggedright\arraybackslash}p{2cm}
  >{\raggedright\arraybackslash}p{3cm}
  >{\raggedright\arraybackslash}p{5cm}
  >{\raggedright\arraybackslash}p{4cm}
  >{\raggedright\arraybackslash}p{2cm}
  >{\raggedright\arraybackslash}p{4cm}
}
\caption{Literature Review on Digital Divide at Individual-level} \label{tab:digitaldivide} \\

\toprule
Year & What is digital divide & Form of digital divide & Factors contributing to digital divide & Channel & Mitigation strategies \\
\midrule
\endfirsthead

\multicolumn{6}{c}{Table \ref{tab:digitaldivide} continued} \\
\toprule
Year & What is digital divide & Form of digital divide & Factors contributing to digital divide & Channel & Mitigation strategies \\
\midrule
\endhead

\midrule
\multicolumn{6}{r}{Continued on next page} \\
\endfoot

\bottomrule
\endlastfoot

Autor et al. 1998 & Not explicitly defined, but focus on skill-biased technological change and its impact on wage inequality & Digital access divide \begin{itemize} \item Access to computers in workplaces \end{itemize} Digital capability divide \begin{itemize} \item IT skills, computer use proficiency \end{itemize} & Education, computer use at work (IT-related), industry capital intensity, industry R\&D expenditures & computer investment, skill upgrading within industries & Not explicitly discussed \\

Jackson et al. 2001 & Differences in Internet use between racial groups & Digital access divide \begin{itemize} \item Access to computers and Internet \end{itemize} Digital capability divide \begin{itemize} \item Computer self-efficacy, Internet use skills, computer anxiety \end{itemize} & Race, gender, age, education, income, cognitive ability, computer self-efficacy, computer anxiety & Internet access, computer ownership & Not explicitly discussed \\

Hargittai 2002 & The gap between those who have access to and use digital technologies and those who do not. & Digital capability divide \begin{itemize} \item Online information-seeking skills \end{itemize} & Age, education level, prior experience with technology, autonomy of use, social support networks & Web use for information retrieval & Education on basic web skills, search strategies, and features of browsers and search engines \\

Van Dijk and Hacker 2003 & Complex and dynamic phenomenon involving unequal access to digital technology & Digital access divide \begin{itemize} \item Material access (possession of computers and network connections) \end{itemize} Digital capability divide \begin{itemize} \item Mental access (motivation and attitudes towards technology use) \item Skills access (digital skills, including online information-seeking skills) \item Usage access (ability to take advantage of usage opportunities) \end{itemize} & Income, education, occupation, age, gender, ethnicity, geographical location & computers, Internet connections & market development of user-friendly ICTs; Tax and income policies for disadvantaged groups; digital skills education; making applications attractive to underrepresented groups \\

Hargittai 2003 & Differences in people's ability to efficiently and effectively find information on the Web & Digital capability divide \begin{itemize} \item Web use efficiency \item Information retrieval skills \end{itemize} & Age, education, presence of children in household, social support networks, experience with the Internet & Web use for information retrieval & Improve search skills, increase time spent online, utilize social support networks for learning \\

Fairlie 2004 & Substantial disparities between racial groups in access to computers and the Internet at home & \begin{minipage}[t]{5cm} Digital access divide \\ Digital capability divide \begin{itemize} \item Computer literacy \end{itemize} \end{minipage} & Race, education, income, occupation, family structure, language barriers & Home computer ownership and Internet access & Improve education and income levels, address language barriers, increase exposure to technology in schools and workplaces \\

Hsieh et al. 2008 & Inequality in access to and use of ICT & \begin{minipage}[t]{5cm} Digital access divide \\ Digital capability divide \begin{itemize} \item ICT self-efficacy \end{itemize} \end{minipage} & Socioeconomic status: age, education, income. Self-efficacy, social influence & Internet TV initiative & Provide free Internet access and equipment \\

Hsieh et al. 2011 & Inequality in access to and use of ICT, particularly among the socioeconomically disadvantaged (SED). & \begin{minipage}[t]{5cm} Digital access divide \\ Digital capability divide \begin{itemize} \item Motivational divide (habitus)
\item ICT knowledge
\item Self-efficacy \end{itemize} \end{minipage} & Socioeconomic status: income, education. Habitus: motivation. Cultural capital: knowledge, self-efficacy. Social capital: Influence from social networks, social support & Free Internet TV provided by local government & Provide free ICT access, Enhance intrinsic motivation and self-efficacy, Leverage social influence and support, Provide training \\

Helsper 2012 & Inequalities in access, skills, use and outcomes of digital technologies & \begin{minipage}[t]{5cm} Digital access divide \\ Digital capability divide \begin{itemize} 
\item Digital skills
\item Usage patterns \end{itemize} \end{minipage} & Economic resources: income, employment. Cultural resources: education level, ethnic background. Social resources: social networks, community support. Personal resources: self-efficacy, health & Internet & Address corresponding offline and online fields of exclusion \\

Venkatesh et al. 2014 & Inequality in access to and use of e-government services & \begin{minipage}[t]{5cm} Digital access divide \\ Digital capability divide \begin{itemize} 
\item e-government portal use skills
\end{itemize} \end{minipage} & Age, gender, education, income, occupation, Personality traits (neuroticism, extraversion, openness to experience, agreeableness, conscientiousness) & E-government portals & Tailor e-government services to meet the needs and preferences of different demographic and personality groups; provide training and support to increase digital literacy and confidence in using e-government services \\

Burtch and Chan 2019 & Inequality in access to and use of medical crowdfunding platforms, leading to differences in financial outcomes such as personal bankruptcy. & Digital capability divide \begin{itemize} \item Digital literacy \item Crowdfunding platform use skills \end{itemize} Digital outcome divide & Socioeconomic status: income, education, occupation. healthcare access, digital literacy & Medical crowdfunding platforms & Improve digital literacy, provide targeted support for vulnerable groups, and ensure equitable access to healthcare resources. \\

Boeing 2020 & Disparities in access to and representation in online rental housing listings & \begin{minipage}[t]{5cm} Digital access divide \\ Digital capability divide \begin{itemize} 
\item The capability to fully exploit online rental platforms
\end{itemize} \end{minipage} & Race, ethnicity, income, education, neighborhood characteristics & Online rental housing listings & Increase transparency and accessibility of online housing platforms \\

Robinson et al. 2020 & Disparities in access to, use of, and outcomes from digital technologies & \begin{minipage}[t]{5cm} Digital access divide \\ Digital capability divide \begin{itemize} 
\item Digital literacy (Digital Literacy: Ability to search for information, evaluate sources, and navigate online platforms)
\item Usage patterns (Differences in how technology is used (e.g., for entertainment vs. educational purposes)
\end{itemize} 
Digital outcome divide
\end{minipage} & Age, gender, education, income, race/ethnicity, disability, geography & Internet and digital devices & Improve education, increase access, develop digital skills \\

Hidalgo et al. 2020 & The unequal access to and use of information and communication technologies (ICT). & Digital capability divide \begin{itemize} \item ICT skills \item Digital competences \end{itemize} & Gender, age, education level, occupation, household income (monthly), habitat (number of inhabitants) & PCs and Internet & Incorporate digital competences in education, build infrastructure, reduce Internet access costs, focus on both national and local government policies \\

Yang and Zhang 2023 & Complex phenomenon involving unequal access to and use of ICT, shaped by user-platform interactions & \begin{minipage}[t]{5cm} Digital access divide \\ Digital capability divide \begin{itemize} 
\item Technology self-efficacy
\item Platform usage skills
\end{itemize} \end{minipage} & Platform affordance (technology-efficacy and self-efficacy), Sociodemographic factors: gender, age, education, occupation, income, Online participation & Social media platforms & Understanding platform affordances, promoting active online participation \\

Celik 2023 & Inequalities in accessing and using various technologies, including AI-based tools & \begin{minipage}[t]{5cm} Digital capability divide \begin{itemize} 
\item AI literacy
\end{itemize} 
Digital outcome divide
\end{minipage} & Education, motivation, physical access, skills access, usage access, computational thinking & AI literacy in online news and entertainment context & Improve education programs, increase access to ICTs, develop computational thinking skills \\

Kohli et al. 2024 & Disparities in access to broadband internet and mental healthcare resources & \begin{minipage}[t]{5cm} Digital access divide \\  Digital outcome divide
\end{minipage} & Urbanization level, poverty rate, geographical location & Broadband internet, telemedicine, mental healthcare facilities and providers & Invest in broadband infrastructure, train rural mental health providers, use mobile phone-based technology, implement integrated care models \\

Haslberger et al. 2023 & Inequality in the use and benefits of AI technology in the labor market, particularly among different demographic groups. & Digital outcome divide (outcomes of AI engagement) & Age, education, occupation, familiarity with technology, gender & AI tools (e.g., ChatGPT) & Provide targeted training and support for older workers and those less familiar with technology \\

Wang et al. 2024 & Disparities in AI-related competencies (knowledge, skills, and attitudes) among users & \begin{minipage}[t]{5cm} Digital capability divide \begin{itemize} 
\item AI Knowledge
\item AI Skills
\item Privacy protection skills
\end{itemize} 
Digital outcome divide (outcomes of AI engagement)
\end{minipage} & Age, education, privacy protection skills, gender & Conventional AI-shaped online news and entertainment environment & Improve education programs, Develop explainable AI, Increase transparency in AI systems \\

\end{longtable}

\scriptsize
\begin{longtable}{
    >{\raggedright\arraybackslash} p{2cm}
    >{\raggedright\arraybackslash} p{2.5cm}
    >{\raggedright\arraybackslash} p{6.5cm}
    >{\raggedright\arraybackslash} p{2cm}
    >{\raggedright\arraybackslash} p{3cm}
    >{\raggedright\arraybackslash} p{2cm}
    >{\raggedright\arraybackslash} p{2cm}}
\caption{Literature Review on AI Adoption} \label{tab:aiadoption} \\

\toprule
Reference & Topic & Focus & Type of AI adoption & AI technology & Level of adoption & Method \\
\midrule
\endfirsthead

\multicolumn{7}{c}{Table \ref{tab:aiadoption} continued} \\
\toprule
Reference & Topic & Focus & Type of AI adoption & AI technology & Level of adoption & Method \\
\midrule
\endhead

\midrule
\multicolumn{7}{r}{Continued on next page} \\
\endfoot

\bottomrule
\endlastfoot

Alekseeva et al. 2020 & What are the outcomes of AI adoption? & Outcomes:
\begin{itemize}
\item Firm growth
\item Productivity
\item Investment decisions
\end{itemize} & Non-generative AI & Machine learning, deep learning, image processing, speech recognition etc. & Organization & Econometrics \\

Gans 2023 & What are the outcomes of AI adoption? & Outcomes:
\begin{itemize}
\item Firm profitability
\item Price and quantity decisions
\item Consumer welfare
\end{itemize} & Non-generative AI & Demand prediction analytics & Organization & Analytical model \\

McElheran et al. 2024 & Who are adopting AI? & Driving factors:
\begin{itemize}
\item Firm size
\item Age
\item Industry
\item Location
\item Owner characteristics
\item Startup financing and innovation strategies
\end{itemize} & Non-generative AI & Machine learning, machine vision, natural language processing, voice recognition, automated guided vehicles etc. & Industry & Survey, Econometrics \\

Agrawal et al. 2024 & What are the outcomes of AI adoption? & Outcomes:
\begin{itemize}
\item Decision variation
\item Alignment challenges
\item System-wide changes
\end{itemize} & Non-generative AI & AI for prediction & Organization & Analytical model \\

Zhang 2024 & What are the outcomes of AI adoption? & Outcomes:
\begin{itemize}
\item Economic dynamics
\item Asset pricing
\end{itemize} & Non-generative AI & Not specified & Organization  & Analytical model \\

Chen et al. 2024 & What are the determinants of AI adoption? & Driving factors:
\begin{itemize}
\item Work mobility
\end{itemize} & Non-generative AI & Machine Learning (ML) in enterprise business analytics software & Organization & Econometrics (two-period long-differences model) \\

Prasad Agrawal 2023 & What are the determinants of generative AI adoption? & Driving factors:
\begin{itemize}
\item Compatibility
\item Complexity
\item Organizational size
\item Regulatory support
\item Environmental uncertainty
\item Competition intensity
\end{itemize} & Generative AI & ChatGPT, GPT-4, Bart, GitHub & Organization & Survey, Econometrics (Logistic Regression) \\

Jo and Bang 2023 & What are the determinants of AI adoption? & Driving Factors:
\begin{itemize}
\item Network quality
\item Accessibility
\item System response
\item Organizational culture
\item Social influence
\item Knowledge application
\end{itemize} & Generative AI & ChatGPT & Individual & Survey + Structural Equation Modeling (SEM) \\

Ma et al. 2024 & What are the determinants of AI adoption? & Driving factors:
\begin{itemize}
\item Perceived usefulness
\item Perceived ease of use
\end{itemize} & Generative AI & ChatGPT & Individual & Survey (Structural Equation Modeling) \\

Russo 2024 & What are the determinants of AI adoption? & Driving factors:
\begin{itemize}
\item Compatibility with existing workflows
\item Perceived usefulness
\item Ease of use
\end{itemize} & Generative AI & Large Language Models (LLMs) & Organization & Survey and PLS-SEM \\

\end{longtable}

\end{landscape}

\newpage
\section{Social Characteristics, AI Literacy, and Generative AI Divide}
\label{appendix:theory}

In this Appendix, we elaborate on the mechanisms through which social characteristics shape heterogeneity in learning and utility from the adoption of generative AI. We focus on how these characteristics influence users’ AI literacy, which governs both belief updating about system capabilities (learning) and value extraction conditional on use (utility).

Education is expected to be a core determinant of AI literacy, shaping both learning and utility outcomes \citep{scheerder2017determinants, dewan2005digital, van2003digital}. Education can enhance users' capacity to assess information quality, reason under uncertainty, and interpret complex socio-technical systems \citep{celik2023exploring, scheerder2017determinants, hargittai2002second}. These competencies are particularly important in the context of generative AI, where output quality is not directly observable from accuracy alone and instead must be inferred from probabilistic feedback accumulated across interactions \citep{xu2024hallucination, yan2024promises}. Therefore, higher education is expected to strengthen AI literacy by enabling users to aggregate signals, discount misleading outputs, and update beliefs about system capability more efficiently \citep{celik2023exploring}. As a result, more educated users might learn system capabilities more rapidly. In addition, better-educated users with higher AI literacy can articulate clearer intentions and frame tasks more precisely \citep{annapureddy2025generative, ng2021conceptualizing}. Because generative AI tends to produce higher-quality outputs when users provide well-structured inputs, higher education is also expected to translate into greater per-use utility \citep{celik2023exploring}.

Age is expected to shape AI literacy through systematic differences in mental operation abilities and technological exposure \citep{scheerder2017determinants, dewan2005digital, van2003digital}. In general, older adults are slower in performing the mental operations needed to integrate new and complex knowledge into long-term memory \citep{salthouse2012consequences}. They also face higher time costs of accumulating new knowledge, which can reduce their exposure to modern digital technologies, especially rapidly evolving AI technologies \citep{lythreatis2022digital, scheerder2017determinants}. Therefore, we expect older individuals to have lower AI literacy, which translates into slower learning dynamics and lower per-use utility \citep{yan2024promises, celik2023exploring}.

Race and English literacy can shape AI literacy in the generative AI context through differences in representational alignment. Because LLM-based generative AI is language-based and trained on large-scale corpora that disproportionately reflect dominant English-language and Western cultural contexts \citep{liu2024datasets, zhang2024regurgitative}, users whose linguistic practices and cultural reference frames are more closely aligned with these dominant representations, in particular, white users and users with higher English literacy, may face systematically different experiences. Such alignment in language and culture can itself be viewed as a form of literacy in the generative AI context: English prompts are more heavily represented in training data and may therefore yield better results, and outputs may also better match white users' preferences through cultural fit \citep{agarwal2025ai, martinez2025llms, qin2025survey}. Therefore, we expect white users and users with higher English literacy to extract higher utility per use and to learn system capability more rapidly.

Gender may be associated with AI literacy because men and women can differ in their pathways of skill accumulation. Prior research documents persistent gender gaps in advanced digital skills, technology experimentation, and participation in technically intensive or AI-adjacent environments \citep{robinson2020digital, scheerder2017determinants, hargittai2002second}. Men are more likely to be situated in formal and informal learning contexts that provide repeated exposure to AI tools, computational thinking, and technology use, which are critical for developing AI literacy \citep{mcelheran2024ai, celik2023exploring}. AI literacy can then facilitate faster belief updating about system capabilities and more effective prompt formulation through sustained interactions and feedback. Accordingly, male users are expected to exhibit faster learning and higher per-use utility in their generative AI use.

LLM occupational exposure (LLMOE) plays a central role in shaping AI literacy by determining who obtains sustained, task-relevant interaction with generative AI \citep{eloundou2024gpts}. Occupational context is a primary channel through which exposure to AI technologies is unevenly distributed, beyond individual access or other social characteristics \citep{mcelheran2024ai, vassilakopoulou2023bridging, dewan2005digital}. Individuals in AI-adjacent or computationally intensive occupations are more likely to encounter AI as part of routine workflows, which facilitates repeated use, feedback-driven learning, and peer-supported knowledge transfer, all of which are important for developing AI literacy \citep{celik2023exploring, scheerder2017determinants}. In the generative AI context, such occupational exposure can accelerate belief updating by improving users' ability to interpret probabilistic outputs, calibrate trust under uncertainty, and generalize from noisy realizations, while also enhancing their capability to formulate effective prompts \citep{yan2024promises, annapureddy2025generative}. As a result, higher levels of LLMOE are expected to be associated with faster learning of system capabilities and higher per-use utility from generative AI interactions.

\clearpage
\section{Summary Statistics for the Full Dataset}
\label{app:summary_stats}
In this appendix, we present the summary statistics for the full sample. 

\begin{table}[h] \centering \footnotesize
  \caption{Summary Statistics (Full Data)} 
  \label{tab:summary_statistics_description} 
\begin{tabular}{@{\extracolsep{5pt}}lp{160pt}rrrrrr} 
\toprule
Variable & \multicolumn{1}{c}{Description} & \multicolumn{1}{c}{N} & \multicolumn{1}{c}{Mean} & \multicolumn{1}{c}{Std} & \multicolumn{1}{c}{Max} & \multicolumn{1}{c}{Min} & \multicolumn{1}{c}{Median} \\ 
\midrule
\textit{College}  & A binary indicator of whether the user receives education after high school. & 11,752 & 0.44 & 0.50 & 1 & 0 & 0 \\
\textit{Age} & The user's age in integers. & 11,752 & 44.82 & 14.46 & 98 & 18 & 43 \\
\textit{Male} & A binary indicator of whether the user is male or not. & 11,752 & 0.38 & 0.48 & 1 & 0 & 0 \\
\textit{White} & A binary indicator of whether the user is a white person. & 11,752 & 0.73 & 0.44 & 1 & 0 & 1 \\
\textit{LLMOE} & LLM Occupational Exposure. & 11,752 & 0.12 & 0.08 & 0.79 & 0 & 0.11\\
\textit{English} & The user's English literacy measured by the portion of English language queries among all the search queries before the launch of ChatGPT. & 11,752 & 0.78 & 0.22 & 1.00 & 0.00 & 0.85\\
\bottomrule
\end{tabular} 
\end{table}

\clearpage
\section{Variable Distribution}
\label{app:variable_distribution}

In this appendix, we report the distributions of all social characteristics for the full sample and for ChatGPT users. For the three binary variables, note that each figure uses two y-axes: the left axis corresponds to the full sample, and the right axis corresponds to ChatGPT users.

\begin{figure}[ht]
    \centering
    \begin{minipage}{0.45\textwidth}
        \includegraphics[width=\textwidth]{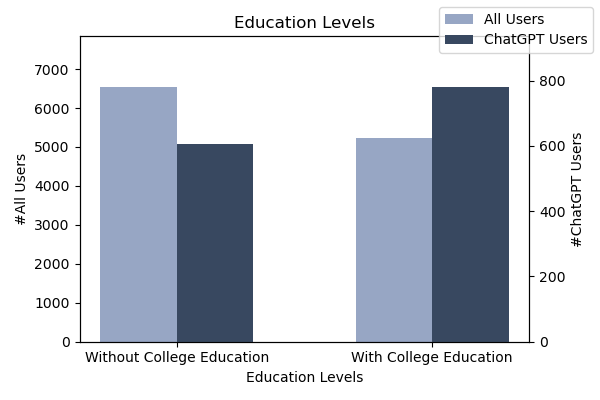}
    \end{minipage}
    \hfill
    \begin{minipage}{0.45\textwidth}
        \includegraphics[width=\textwidth]{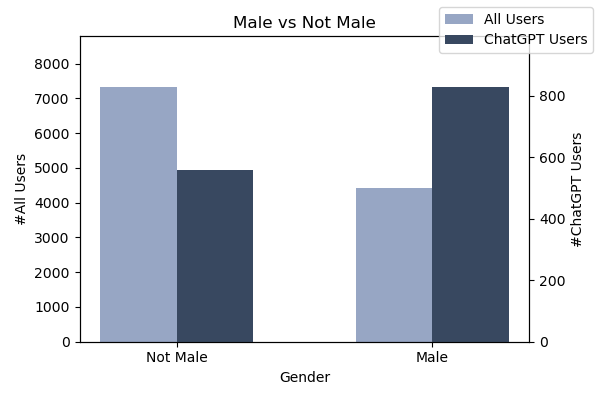}
    \end{minipage}
    \vspace{0.5cm}
    
    \begin{minipage}{0.45\textwidth}
        \includegraphics[width=\textwidth]{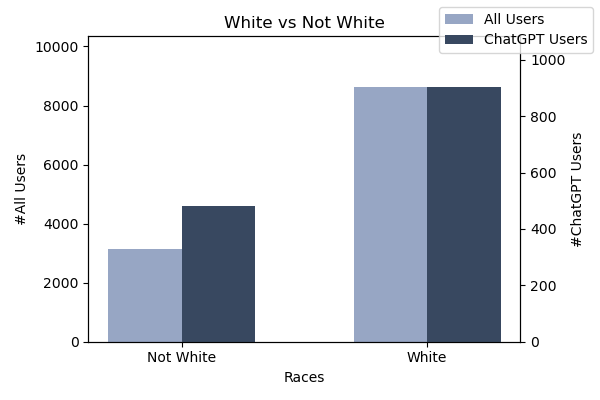}
    \end{minipage}
    \hfill
    \begin{minipage}{0.45\textwidth}
        \includegraphics[width=\textwidth]{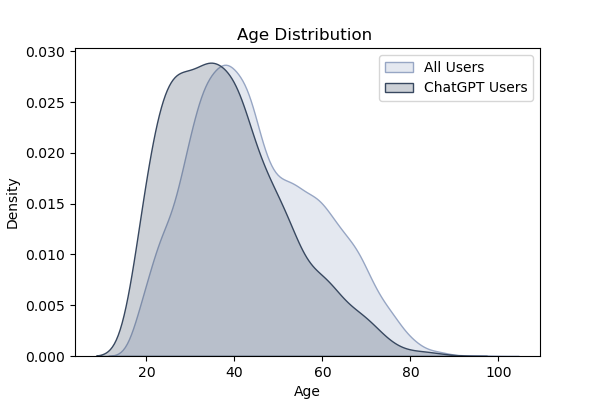}
    \end{minipage}

    \vspace{0.5cm}

    \begin{minipage}{0.45\textwidth}
        \centering
        \includegraphics[width=\textwidth]{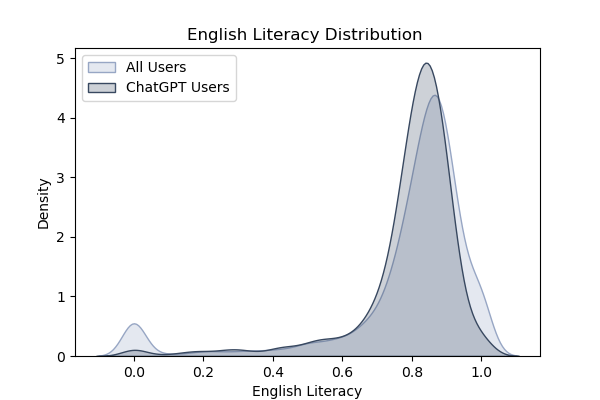}
    \end{minipage}
    \hfill
    \begin{minipage}{0.45\textwidth}
        \includegraphics[width=\textwidth]{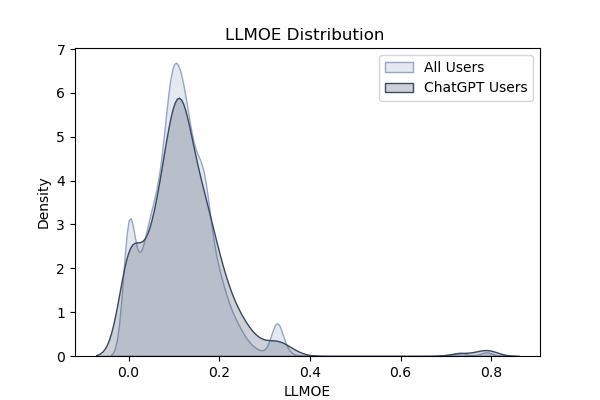}
    \end{minipage}

    \caption{Demographic distributions for all users and ChatGPT users.}
    \label{fig:demo}
\end{figure}

\newpage
\section{Comparison with Population Distribution}
\label{app:comparison_with_population}

Here we compare the distribution of demographic features in our sample with the U.S. population to show the representativeness of our data. The distribution of $College$, $Age$, $Male$, $White$, $LLMOE$, and $English$ are from the United States Census Bureau\footnote{See https://www.census.gov/data.html}. The $LLMOE$ is from \cite{eloundou2024gpts}.

\begin{table}[h]
\centering \small
\caption{Distribution Comparison: Our Dataset vs.\ US Population}
\label{tab:demo_vs_population}
\begin{tabular}{lcc}
\toprule
Demographic Variable & Our dataset (Mean) & US population (Mean) \\
\midrule
\textit{College}            & 0.44 & 0.62 \\
\textit{Age}                & 44.82 & 39.81 \\
\textit{Male}               & 0.38 & 0.48 \\
\textit{White}              & 0.73 & 0.75 \\
\textit{LLMOE}               & 0.12 & 0.14 \\
\textit{English}            & 0.78 & 0.78 \\
\bottomrule
\end{tabular}

\end{table}

\clearpage
\section{Correlation Matrix}
\label{app:corr_matrix}

Here we present the correlation matrix for our focal demographic features: $College$, $Age$, $Male$, $White$, $LLMOE$, and $English$. Overall, the correlations among these variables are not significant.

\begin{figure}[h]
\centering
    \begin{minipage}{0.7\textwidth}
        \centering
        \includegraphics[width=\textwidth]{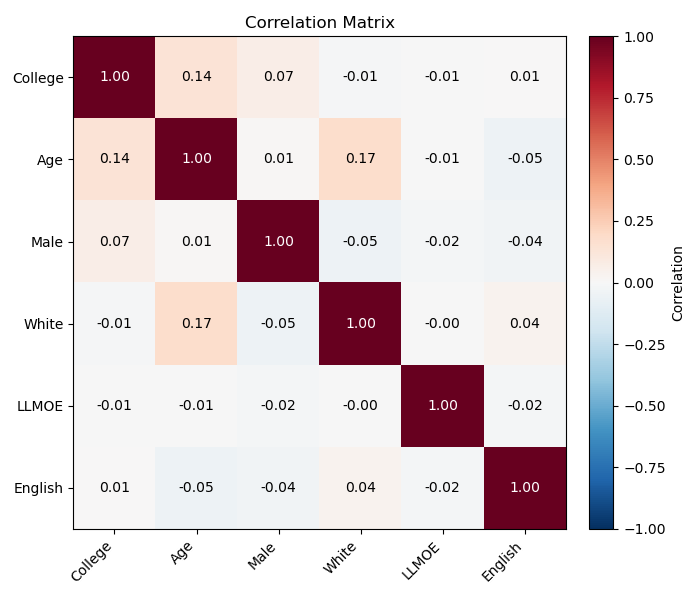}
    \end{minipage}
    \label{fig:correlation_matrix}
    \caption{Correlation Matrix}
\end{figure}

\clearpage
\section{Reduced-Form Analyses}
\label{app:reduce_form}
In this appendix, we examine how users’ average daily ChatGPT usage frequency relates to demographic characteristics within the subset of ChatGPT users. We construct $GPTFrequency$ to measure the daily frequency of interactions with ChatGPT per day after the users use ChatGPT for the first time. We find that users with some college education, younger users, males, white users, and those with higher LLM occupational exposure and greater English literacy use ChatGPT more frequently.

\begin{table}[htbp] \centering \small
  \caption{Estimation Results} 
  \label{tab:results_reduce_form} 
    \begin{tabular}{lc} 
    \toprule
 & (1) \\
Variables & \textit{GPTFrequency}  \\ 
\midrule
\textit{College} & 0.188$^{**}$   \\
 & (0.091)   \\
\textit{AgeHigh} & $-$0.5596$^{***}$   \\
 & (0.193)  \\
\textit{AgeMid} & $-$0.1885$^{**}$   \\
 & (0.088)  \\
\textit{Male} & 0.1369$^{*}$   \\
 & (0.081)  \\
\textit{White} & $-$0.6050$^{***}$   \\
 & (0.215)  \\
\textit{LLMOE} & 2.117$^{***}$ \\
 & (0.698) \\
\textit{English} & $-$0.5532$^{**}$ \\
 & (0.225) \\
Constant & 1.907$^{***}$  \\
 & (0.668)  \\
\midrule
 Observations & 1,344  \\ 
\bottomrule
\multicolumn{2}{l}{ \textit{Note}: Standard errors in parentheses; $^{***}$ p$<$0.01, $^{**}$ p$<$0.05, $^{*}$ p$<$0.1.} \\
\end{tabular}
\end{table}

\clearpage
\section{Model-Free Evidence on ChatGPT and Search Engine Usage}

We provide model-free evidence on weekly average search and ChatGPT usage by week. Here, we focus on the subset of users who use ChatGPT in each of eight consecutive weeks, so that the entire adoption process can be observed. We observe that as users’ ChatGPT usage increases, their search engine usage declines.

\begin{figure}[h]
\centering
    \begin{minipage}{0.9\textwidth}
        \centering
        \includegraphics[width=\textwidth]{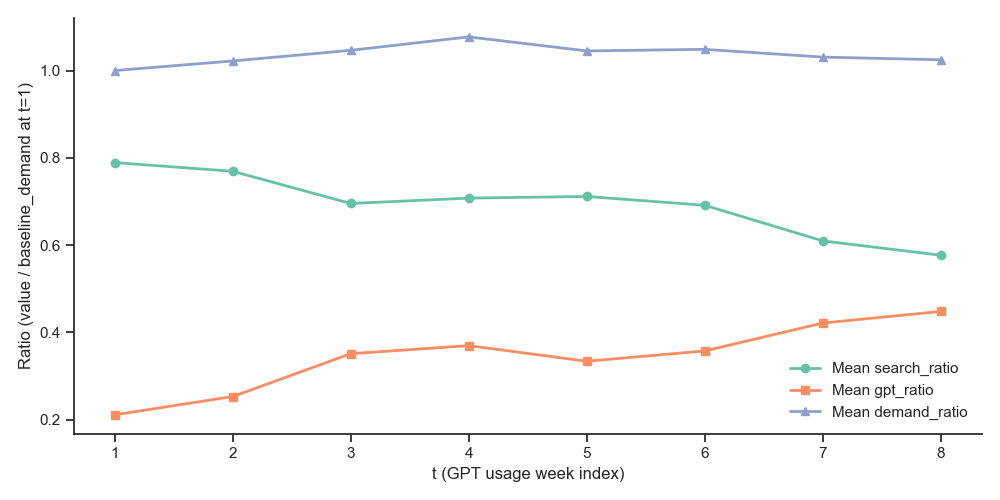}
        \caption{Weekly Average Search and ChatGPT Usage by Week}
        \label{fig:model_free}
    \end{minipage}
    
\end{figure}

\clearpage
\section{Model Selection}
\label{app:model_select}

As a robustness check, we estimate models with one to four latent classes and report the corresponding BIC values. The results show a substantial decline in BIC when moving from the no-latent-class specification to the two-class specification. By contrast, adding further classes yields only modest changes in BIC. Therefore, we adopt the two-class model as our main specification \citep{huang2019level}.

\begin{table}[!htbp]
\centering
\caption{Model Fit Comparison Across Latent-Class Specifications}
\label{tab:lc_model_fit}
\begin{tabular}{lcc}
\toprule
Model (Number of Latent Classes) & Log-likelihood & BIC \\
\midrule
model with no latent class & $-141,614$ & $283,358$ \\
2-class model (our main model) & $-139,964$ & $280,086$ \\
3-class model & $-139,873$ & $279,933$ \\
4-class model & $-139,810$ & $279,836$ \\
\bottomrule
\end{tabular}
\end{table}

\clearpage
\section{Robustness Check for Data Before the Launch of GPT-4}
\label{app:robustness_gpt4}

In this robustness check, we aim to rule out the potential confounding effect of ChatGPT’s major upgrade from GPT-3 to GPT-4. We restrict the sample to observations prior to the GPT-4 launch on March 14, 2023, and reestimate the model. The results are reported in Table \ref{tab:results_gpt4}.

\begin{table}[htbp] \footnotesize
\centering
\caption{Parameters Estimation for Data Before the Launch of GPT-4}
\label{tab:results_gpt4}
\resizebox{0.8\textwidth}{!}{%
\begin{tabular}{l p{11cm} c c }
\toprule
\textbf{Parameter} & \textbf{Description} & \textbf{Value} & \textbf{Std} \\
\midrule
$v_0$ & Prior belief on the mean value of representative utility & 0 & -- (fixed) \\
log($\sigma_0^2$) & Prior belief on the variance of representative utility & 4 & -- (fixed) \\
$\lambda$ & Latent class probability parameter & $-$1.388 & $(0.071)^{***}$ \\
$c$ & Mean representative utility of traditional search engines & 1.034 & $(0.012)^{***}$ \\
$\Delta c$ & Difference in mean representative utility of traditional search engines betwen two latent classes & 0.472 & $(0.015)^{***}$ \\
$\alpha_0$ & Intercept of ChatGPT's representative utility & 0.382 & $(0.054)^{***}$ \\
$log(\Delta\alpha_0)$ & Difference in ChatGPT's representative utility between two latent classes & 0.597 & $(0.020)^{***}$ \\
$\alpha_1(HighEdu)$ & Coefficient of high education levels on ChatGPT's representative utility & $-$0.525 & $(0.014)^{***}$ \\
$\alpha_2(AgeHigh)$ & Coefficient of high age group on ChatGPT's representative utility & $-$1.215 & $(0.023)^{***}$ \\
$\alpha_3(AgeMid)$ & Coefficient of middle age group on ChatGPT's representative utility & $-$0.917 & $(0.016)^{***}$ \\
$\alpha_4(Male)$ & Coefficient of males on ChatGPT's representative utility & 1.446 & $(0.018)^{***}$ \\
$\alpha_5(White)$ & Coefficient of race (white vs non-white) on ChatGPT's representative utility & $-$1.489 & $(0.013)^{***}$ \\
$\alpha_6(LLMOE)$ & Coefficient of LLM exposure on ChatGPT's representative utility  & 1.257 & $(0.032)^{***}$ \\
$\alpha_7(English)$ & Coefficient of English literacy on ChatGPT's representative utility  & $-$1.082 & $(0.058)^{***}$ \\
$\sigma_n^2$ & News signal variance & 5.790 & $(0.021)^{***}$ \\
$\gamma_0$ & Intercept of usage signal variance & 5.325 & $(0.057)^{***}$ \\
$\Delta\gamma_0$ & Difference in usage signal variance between two latent classes & 3.909 & $(0.016)^{***}$ \\
$\gamma_1(HighEdu)$ & Coefficient of high education levels on usage signal variance & $-$0.139 & $(0.015)^{***}$ \\
$\gamma_2(AgeHigh)$ & Coefficient of high age group on usage signal variance & 0.287 & $(0.021)^{***}$ \\
$\gamma_3(AgeMid)$ & Coefficient of middle age group on usage signal variance & 0.195 & $(0.018)^{***}$ \\
$\gamma_4(Male)$ & Coefficient of males on usage signal va
riance& $-$0.216 & $(0.017)^{***}$ \\
$\gamma_5(White)$ & Coefficient of white people on usage signal variance & $-$0.701 & $(0.015)^{***}$ \\
$\gamma_6(LLMOE)$ & Coefficient of LLM exposure on usage signal variance & $-$0.813 & $(0.081)^{***}$ \\
$\gamma_7(English)$ & Coefficient of English literacy on usage signal variance & $-$1.688 & $(0.064)^{***}$ \\
\bottomrule
\multicolumn{3}{l}{ \textit{Note}: Standard errors in parentheses; $^{***}$ p$<$0.01, $^{**}$ p$<$0.05, $^{*}$ p$<$0.1.} \\
\end{tabular}}
\end{table}

\clearpage
\section{Robustness Check for Alternative Specification of Age}
\label{app:robustness_age_2}
To incorporate potential nonlinear effects of age, we alternatively include $Age$ and $Age^2$ in the model. The results indicate a nonlinear relationship between users' representative utility from ChatGPT and $Age$: as age increases, utility declines at a decreasing rate. We obtain a similar pattern for learning speed, which also declines at a decreasing rate with age. These findings are consistent with our main results.

\begin{table}[htbp] \footnotesize
\centering
\caption{Parameters Estimation for Alternative Specification of Age}
\label{tab:results_alternative_age}
\resizebox{0.9\textwidth}{!}{%
\begin{tabular}{l p{11cm} c c }
\toprule
\textbf{Parameter} & \textbf{Description} & \textbf{Value} & \textbf{Std} \\
\midrule
$v_0$ & Prior belief on the mean value of representative utility & 0 & -- (fixed) \\
log($\sigma_0^2$) & Prior belief on the variance of representative utility & 4 & -- (fixed) \\
$\lambda$ & Latent class probability parameter & $-$1.384 & $(0.068)^{***}$ \\
$c$ & Mean representative utility of traditional search engines & 0.930 & $(0.012)^{***}$ \\
$\Delta c$ & Difference in mean representative utility of traditional search engines betwen two latent classes & 0.620 & $(0.013)^{***}$ \\
$\alpha_0$ & Intercept of ChatGPT's representative utility & 0.402 & $(0.067)^{***}$ \\
$log(\Delta\alpha_0)$ & Difference in ChatGPT's representative utility between two latent classes & 0.498 & $(0.012)^{***}$ \\
$\alpha_1(HighEdu)$ & Coefficient of high education levels on ChatGPT's representative utility & $-$0.232 & $(0.052)^{***}$ \\
$\alpha_2(Age)$ & Coefficient of age on ChatGPT's representative utility & $-$0.028 & $(0.004)^{***}$ \\
$\alpha_3(Age^2)$ & Coefficient of age square on ChatGPT's representative utility & 0.0002 & $(0.00005)^{***}$ \\
$\alpha_4(Male)$ & Coefficient of males on ChatGPT's representative utility & 1.474 & $(0.0126)^{***}$ \\
$\alpha_5(White)$ & Coefficient of race (white vs non-white) on ChatGPT's representative utility & $-$1.640 & $(0.014)^{***}$ \\
$\alpha_6(LLMOE)$ & Coefficient of LLM exposure on ChatGPT's representative utility  & 1.952 & $(0.031)^{***}$ \\
$\alpha_7(English)$ & Coefficient of English literacy on ChatGPT's representative utility  & $-$1.517& $(0.091)^{***}$ \\
$\sigma_n^2$ & News signal variance & 5.587 & $(0.023)^{***}$ \\
$\gamma_0$ & Intercept of usage signal variance & 4.090 & $(0.107)^{***}$ \\
$\Delta\gamma_0$ & Difference in usage signal variance between two latent classes & 3.187 & $(0.016)^{***}$ \\
$\gamma_1(HighEdu)$ & Coefficient of high education levels on usage signal variance & $-$0.024 & $(0.038)^{***}$ \\
$\gamma_2(Age)$ & Coefficient of age on usage signal variance & 0.018 & $(0.003)^{***}$ \\
$\gamma_3(Age^2)$ & Coefficient of age square on usage signal variance & $-$0.0001 & $(0.00002)^{***}$ \\
$\gamma_4(Male)$ & Coefficient of males on usage signal variance& $-$0.049 & $(0.016)^{***}$ \\
$\gamma_5(White)$ & Coefficient of white people on usage signal variance & $-$0.792 & $(0.041)^{***}$ \\
$\gamma_6(LLMOE)$ & Coefficient of LLM exposure on usage signal variance & $-$0.763 & $(0.177)^{***}$ \\
$\gamma_7(English)$ & Coefficient of English literacy on usage signal variance & $-$-0.381 & $(0.05)^{***}$ \\
\bottomrule
\multicolumn{3}{l}{ \textit{Note}: Standard errors in parentheses; $^{***}$ p$<$0.01, $^{**}$ p$<$0.05, $^{*}$ p$<$0.1.} \\
\end{tabular}}
\end{table}

\clearpage
\section{Robustness Check for Alternative Value of Prior Belief Variance}
\label{app:robustness_variance}

In this robustness check, we aim to test the robustness of our model with different prior belief variance. We set the prior belief on the variance of representative utility to $exp(5)$, and reestimate the model. The results are reported in Table \ref{tab:results_alternative_variance}.

\begin{table}[htbp] \footnotesize
\centering
\caption{Parameters Estimation for Alternative Value of Prior Belief Variance}
\label{tab:results_alternative_variance}
\resizebox{0.8\textwidth}{!}{%
\begin{tabular}{l p{11cm} c c }
\toprule
\textbf{Parameter} & \textbf{Description} & \textbf{Value} & \textbf{Std} \\
\midrule
$v_0$ & Prior belief on the mean value of representative utility & 0 & -- (fixed) \\
log($\sigma_0^2$) & Prior belief on the variance of representative utility & 5 & -- (fixed) \\
$\lambda$ & Latent class probability parameter & $-$1.314 & $(0.070)^{***}$ \\
$c$ & Mean representative utility of traditional search engines & 0.946 & $(0.014)^{***}$ \\
$\Delta c$ & Difference in mean representative utility of traditional search engines betwen two latent classes & 0.572 & $(0.016)^{***}$ \\
$\alpha_0$ & Intercept of ChatGPT's representative utility & 0.310 & $(0.050)^{***}$ \\
$log(\Delta\alpha_0)$ & Difference in ChatGPT's representative utility between two latent classes & 0.418 & $(0.020)^{***}$ \\
$\alpha_1(HighEdu)$ & Coefficient of high education levels on ChatGPT's representative utility & $-$0.180 & $(0.012)^{***}$ \\
$\alpha_2(AgeHigh)$ & Coefficient of high age group on ChatGPT's representative utility & $-$1.451 & $(0.019)^{***}$ \\
$\alpha_3(AgeMid)$ & Coefficient of middle age group on ChatGPT's representative utility & $-$1.102 & $(0.014)^{***}$ \\
$\alpha_4(Male)$ & Coefficient of males on ChatGPT's representative utility & 1.003 & $(0.014)^{***}$ \\
$\alpha_5(White)$ & Coefficient of race (white vs non-white) on ChatGPT's representative utility & $-$1.217 & $(0.012)^{***}$ \\
$\alpha_6(LLMOE)$ & Coefficient of LLM exposure on ChatGPT's representative utility  & 1.104 & $(0.029)^{***}$ \\
$\alpha_7(English)$ & Coefficient of English literacy on ChatGPT's representative utility  & $-$0.291& $(0.052)^{***}$ \\
$\sigma_n^2$ & News signal variance & 4.585 & $(0.015)^{***}$ \\
$\gamma_0$ & Intercept of usage signal variance & 4.909 & $(0.055)^{***}$ \\
$\Delta\gamma_0$ & Difference in usage signal variance between two latent classes & 8.183 & $(0.015)^{***}$ \\
$\gamma_1(HighEdu)$ & Coefficient of high education levels on usage signal variance & $-$0.180 & $(0.015)^{***}$ \\
$\gamma_2(AgeHigh)$ & Coefficient of high age group on usage signal variance & 0.211 & $(0.025)^{***}$ \\
$\gamma_3(AgeMid)$ & Coefficient of middle age group on usage signal variance & 0.132 & $(0.016)^{***}$ \\
$\gamma_4(Male)$ & Coefficient of males on usage signal va
riance& $-$0.058 & $(0.016)^{***}$ \\
$\gamma_5(White)$ & Coefficient of white people on usage signal variance & $-$0.462 & $(0.015)^{***}$ \\
$\gamma_6(LLMOE)$ & Coefficient of LLM exposure on usage signal variance & $-$0.547 & $(0.077)^{***}$ \\
$\gamma_7(English)$ & Coefficient of English literacy on usage signal variance & $-$1.603 & $(0.061)^{***}$ \\
\bottomrule
\multicolumn{3}{l}{ \textit{Note}: Standard errors in parentheses; $^{***}$ p$<$0.01, $^{**}$ p$<$0.05, $^{*}$ p$<$0.1.} \\
\end{tabular}}
\end{table}

\clearpage
\section{Robustness Check for a Different List of News Outlets}
\label{app:robustness_alternative_news}

We retrieved the news outlets from the same source\footnote{ \url{https://pressgazette.co.uk/media-audience-and-business-data/media_metrics/most-popular-websites-news-us-monthly-3/}} but at Oct. 2024 instead. We reestimate our model and the results can be found at Table \ref{tab:results_alternative_news_outlets}.

\begin{table}[htbp] \footnotesize
\centering
\caption{Parameters Estimation forDifferent List of News Outlets}
\label{tab:results_alternative_news_outlets}
\resizebox{0.8\textwidth}{!}{%
\begin{tabular}{l p{11cm} c c }
\toprule
\textbf{Parameter} & \textbf{Description} & \textbf{Value} & \textbf{Std} \\
\midrule
$v_0$ & Prior belief on the mean value of representative utility & 0 & -- (fixed) \\
log($\sigma_0^2$) & Prior belief on the variance of representative utility & 4 & -- (fixed) \\
$\lambda$ & Latent class probability parameter & $-$1.415 & $(0.072)^{***}$ \\
$c$ & Mean representative utility of traditional search engines & 1.034 & $(0.013)^{***}$ \\
$\Delta c$ & Difference in mean representative utility of traditional search engines between two latent classes & 0.536 & $(0.016)^{***}$ \\
$\alpha_0$ & Intercept of ChatGPT's representative utility & 0.290 & $(0.056)^{***}$ \\
$log(\Delta\alpha_0)$ & Difference in ChatGPT's representative utility between two latent classes & 0.815 & $(0.016)^{***}$ \\
$\alpha_1(HighEdu)$ & Coefficient of high education levels on ChatGPT's representative utility & $-$0.477 & $(0.015)^{***}$ \\
$\alpha_2(AgeHigh)$ & Coefficient of high age group on ChatGPT's representative utility & $-$1.236 & $(0.022)^{***}$ \\
$\alpha_3(AgeMid)$ & Coefficient of middle age group on ChatGPT's representative utility & $-$0.864 & $(0.016)^{***}$ \\
$\alpha_4(Male)$ & Coefficient of males on ChatGPT's representative utility & 1.434 & $(0.017)^{***}$ \\
$\alpha_5(White)$ & Coefficient of race (white vs non-white) on ChatGPT's representative utility & $-$1.521 & $(0.013)^{***}$ \\
$\alpha_6(LLMOE)$ & Coefficient of LLM exposure on ChatGPT's representative utility  & 1.226 & $(0.031)^{***}$ \\
$\alpha_7(English)$ & Coefficient of English literacy on ChatGPT's representative utility  & $-$0.980 & $(0.060)^{***}$ \\
$\sigma_n^2$ & News signal variance & 5.507 & $(0.020)^{***}$ \\
$\gamma_0$ & Intercept of usage signal variance & 5.385 & $(0.056)^{***}$ \\
$\Delta\gamma_0$ & Difference in usage signal variance between two latent classes & 5.848 & $(0.016)^{***}$ \\
$\gamma_1(HighEdu)$ & Coefficient of high education levels on usage signal variance & $-$0.107 & $(0.015)^{***}$ \\
$\gamma_2(AgeHigh)$ & Coefficient of high age group on usage signal variance & 0.294 & $(0.021)^{***}$ \\
$\gamma_3(AgeMid)$ & Coefficient of middle age group on usage signal variance & 0.187 & $(0.018)^{***}$ \\
$\gamma_4(Male)$ & Coefficient of males on usage signal va
riance& $-$0.172 & $(0.016)^{***}$ \\
$\gamma_5(White)$ & Coefficient of white people on usage signal variance & $-$0.763 & $(0.015)^{***}$ \\
$\gamma_6(LLMOE)$ & Coefficient of LLM exposure on usage signal variance & $-$0.723 & $(0.077)^{***}$ \\
$\gamma_7(English)$ & Coefficient of English literacy on usage signal variance & $-$1.583 & $(0.062)^{***}$ \\
\bottomrule
\multicolumn{3}{l}{ \textit{Note}: Standard errors in parentheses; $^{***}$ p$<$0.01, $^{**}$ p$<$0.05, $^{*}$ p$<$0.1.} \\
\end{tabular}}
\end{table}

\clearpage
\section{Robustness Check for Excluding Outliers}
\label{app:robustness_excluding_outliers}

We remove users with the top 1\% of activity (Search + ChatGPT) during the observation period and re-estimate our model. The results are in Table \ref{tab:results_no_outliers}. 

\begin{table}[htbp] \footnotesize
\centering
\caption{Parameters Estimation for Excluding Outliers}
\label{tab:results_no_outliers}
\resizebox{0.8\textwidth}{!}{%
\begin{tabular}{l p{11cm} c c }
\toprule
\textbf{Parameter} & \textbf{Description} & \textbf{Value} & \textbf{Std} \\
\midrule
$v_0$ & Prior belief on the mean value of representative utility & 0 & -- (fixed) \\
log($\sigma_0^2$) & Prior belief on the variance of representative utility & 4 & -- (fixed) \\
$\lambda$ & Latent class probability parameter & $-$1.241 & $(0.070)^{***}$ \\
$c$ & Mean representative utility of traditional search engines & 0.856 & $(0.014)^{***}$ \\
$\Delta c$ & Difference in mean representative utility of traditional search engines betwen two latent classes & 0.475 & $(0.016)^{***}$ \\
$\alpha_0$ & Intercept of ChatGPT's representative utility & 0.352 & $(0.057)^{***}$ \\
$log(\Delta\alpha_0)$ & Difference in ChatGPT's representative utility between two latent classes & 0.420 & $(0.043)^{***}$ \\
$\alpha_1(HighEdu)$ & Coefficient of high education levels on ChatGPT's representative utility & $-$0.325 & $(0.013)^{***}$ \\
$\alpha_2(AgeHigh)$ & Coefficient of high age group on ChatGPT's representative utility & $-$1.162 & $(0.023)^{***}$ \\
$\alpha_3(AgeMid)$ & Coefficient of middle age group on ChatGPT's representative utility & $-$0.825 & $(0.020)^{***}$ \\
$\alpha_4(Male)$ & Coefficient of males on ChatGPT's representative utility & 0.886 & $(0.020)^{***}$ \\
$\alpha_5(White)$ & Coefficient of race (white vs non-white) on ChatGPT's representative utility & $-$0.589 & $(0.013)^{***}$ \\
$\alpha_6(LLMOE)$ & Coefficient of LLM exposure on ChatGPT's representative utility  & 2.238 & $(0.025)^{***}$ \\
$\alpha_7(English)$ & Coefficient of English literacy on ChatGPT's representative utility  & $-$2.221 & $(0.061)^{***}$ \\
$\sigma_n^2$ & News signal variance & 4.641 & $(0.024)^{***}$ \\
$\gamma_0$ & Intercept of usage signal variance & 4.591 & $(0.063)^{***}$ \\
$\Delta\gamma_0$ & Difference in usage signal variance between two latent classes & 6.180 & $(0.019)^{***}$ \\
$\gamma_1(HighEdu)$ & Coefficient of high education levels on usage signal variance & $-$0.199 & $(0.018)^{***}$ \\
$\gamma_2(AgeHigh)$ & Coefficient of high age group on usage signal variance & 0.663 & $(0.022)^{***}$ \\
$\gamma_3(AgeMid)$ & Coefficient of middle age group on usage signal variance & 0.609 & $(0.026)^{***}$ \\
$\gamma_4(Male)$ & Coefficient of males on usage signal variance& $-$0.046 & $(0.021)^{***}$ \\
$\gamma_5(White)$ & Coefficient of white people on usage signal variance & $-$0.107 & $(0.018)^{***}$ \\
$\gamma_6(LLMOE)$ & Coefficient of LLM exposure on usage signal variance & $-$0.674 & $(0.056)^{***}$ \\
$\gamma_7(English)$ & Coefficient of English literacy on usage signal variance & $-$1.392 & $(0.069)^{***}$ \\
\bottomrule
\multicolumn{3}{l}{ \textit{Note}: Standard errors in parentheses; $^{***}$ p$<$0.01, $^{**}$ p$<$0.05, $^{*}$ p$<$0.1.} \\
\end{tabular}}
\end{table}

\clearpage
\section{Robustness Check on the Substitution Effect}
\label{app:robustness_substitution}

In this appendix, we conduct a robustness check to more precisely examine the relationship between substitutable search demand (queries) and ChatGPT usage. Specifically, we use the latest OpenAI API to classify users’ queries into informational, navigational, and transactional categories \citep{broder2002taxonomy}. Among these, we expect the clearest substitution to occur between informational queries and ChatGPT usage. We therefore exclude navigational and transactional queries and re-estimate the model using only informational queries. The results remain robust under this specification.

\begin{table}[htbp] \footnotesize
\centering
\caption{Parameters Estimation for Robustness Check on the Substitution Effect}
\label{tab:results_sub}
\resizebox{0.8\textwidth}{!}{%
\begin{tabular}{l p{11cm} c c }
\toprule
\textbf{Parameter} & \textbf{Description} & \textbf{Value} & \textbf{Std} \\
\midrule
$v_0$ & Prior belief on the mean value of representative utility & 0 & -- (fixed) \\
log($\sigma_0^2$) & Prior belief on the variance of representative utility & 4 & -- (fixed) \\
$\lambda$ & Latent class probability parameter & $-$1.244 & $(0.069)^{***}$ \\
$c$ & Mean representative utility of traditional search engines & 1.000 & $(0.013)^{***}$ \\
$\Delta c$ & Difference in mean representative utility of traditional search engines betwen two latent classes & 0.503 & $(0.016)^{***}$ \\
$\alpha_0$ & Intercept of ChatGPT's representative utility & 0.400 & $(0.054)^{***}$ \\
$log(\Delta\alpha_0)$ & Difference in ChatGPT's representative utility between two latent classes & 0.213 & $(0.027)^{***}$ \\
$\alpha_1(HighEdu)$ & Coefficient of high education levels on ChatGPT's representative utility & $-$0.412 & $(0.015)^{***}$ \\
$\alpha_2(AgeHigh)$ & Coefficient of high age group on ChatGPT's representative utility & $-$1.523 & $(0.025)^{***}$ \\
$\alpha_3(AgeMid)$ & Coefficient of middle age group on ChatGPT's representative utility & $-$1.189 & $(0.018)^{***}$ \\
$\alpha_4(Male)$ & Coefficient of males on ChatGPT's representative utility & 1.695 & $(0.019)^{***}$ \\
$\alpha_5(White)$ & Coefficient of race (white vs non-white) on ChatGPT's representative utility & $-$1.105 & $(0.014)^{***}$ \\
$\alpha_6(LLMOE)$ & Coefficient of LLM exposure on ChatGPT's representative utility  & 1.181 & $(0.045)^{***}$ \\
$\alpha_7(English)$ & Coefficient of English literacy on ChatGPT's representative utility  & $-$1.538 & $(0.058)^{***}$ \\
$\sigma_n^2$ & News signal variance & 5.899 & $(0.021)^{***}$ \\
$\gamma_0$ & Intercept of usage signal variance & 5.662 & $(0.055)^{***}$ \\
$\Delta\gamma_0$ & Difference in usage signal variance between two latent classes & 3.384 & $(0.016)^{***}$ \\
$\gamma_1(HighEdu)$ & Coefficient of high education levels on usage signal variance & $-$0.476 & $(0.016)^{***}$ \\
$\gamma_2(AgeHigh)$ & Coefficient of high age group on usage signal variance & 0.503 & $(0.022)^{***}$ \\
$\gamma_3(AgeMid)$ & Coefficient of middle age group on usage signal variance & 0.257 & $(0.018)^{***}$ \\
$\gamma_4(Male)$ & Coefficient of males on usage signal variance& $-$0.144 & $(0.017)^{***}$ \\
$\gamma_5(White)$ & Coefficient of white people on usage signal variance & $-$0.623 & $(0.015)^{***}$ \\
$\gamma_6(LLMOE)$ & Coefficient of LLM exposure on usage signal variance & $-$0.501 & $(0.105)^{***}$ \\
$\gamma_7(English)$ & Coefficient of English literacy on usage signal variance & $-$1.242 & $(0.068)^{***}$ \\
\bottomrule
\multicolumn{3}{l}{ \textit{Note}: Standard errors in parentheses; $^{***}$ p$<$0.01, $^{**}$ p$<$0.05, $^{*}$ p$<$0.1.} \\
\end{tabular}}
\end{table}

\clearpage
\section{Counterfactual Analysis}
\label{app:counterfactual}

In this appendix, we present the figures from the counterfactual simulations described in Section~\ref{sec:policy_simulation}. We focus on two representative users mentioned in Section~\ref{sec:policy_simulation}: one fast learner and one slow learner. Figure~\ref{fig:scatter} plots a single simulated belief path for the slow learner (red dots) and illustrates the emergence of a long-run belief-trap dynamic. We then compare trap rates for the fast and slow learners by running 10{,}000 simulations for each and plotting the resulting error bars. As shown in Figure~\ref{fig:error_bar}, the two users differ substantially in their probabilities of becoming trapped. Importantly, this gap is not primarily driven by differences in utility: when we counterfactually set the slow learner's true utility equal to that of the fast learner (third bar in Figure~\ref{fig:error_bar}), the slow learner's trap rate remains far above that of the fast learner.

A natural follow-up question is how to mitigate the impact of the learning divide. We consider two main approaches based on our model estimates. The first is learning by training, in which users are trained to use ChatGPT for a certain number of rounds to update their beliefs. The second is learning by reading, in which users update their beliefs by reading relevant news information.

As shown in the last two bars of Figure~\ref{fig:error_bar}, providing training at the outset allows slow learners to attain trap rates comparable to those of fast learners after roughly 200 to 400 instances of ChatGPT use, which corresponds to fewer than 25 days of information-seeking activities on average. For comparison, we conduct additional policy simulations to assess how long it would take slow learners to reach a similar trap rate if, instead, they were allowed to read AI-related news every day at the average frequency observed in our sample (Figure~\ref{fig:error_bar_news}). The results indicate that it would require approximately 2 to 4 years of news exposure for slow learners to approach the trap rate of fast learners. Therefore, training appears to be a more natural, feasible, and cost-effective approach to help users learn ChatGPT's utility and mitigate the belief trap.

Finally, we examine how future upgrades to generative AI products might help the belief-trap issue. For an improved version of ChatGPT, we consider two counterfactual scenarios: (1) the new version delivers higher utility to users, and (2) the new version enables users to learn faster. We conduct separate analyses for each case. The simulations show that a slow learner attains a trap rate comparable to that of the original fast learner only if the upgraded ChatGPT increases the slow learner's utility by roughly 300\% to 1000\% or raises learning speed by approximately 10 to 50 times. These results suggest that improvements in LLM tools can facilitate adoption, but the required performance gains are likely to be quite substantial.

\begin{figure}[ht]
  \centering

  \begin{minipage}{0.7\textwidth}
    \includegraphics[width=\textwidth]{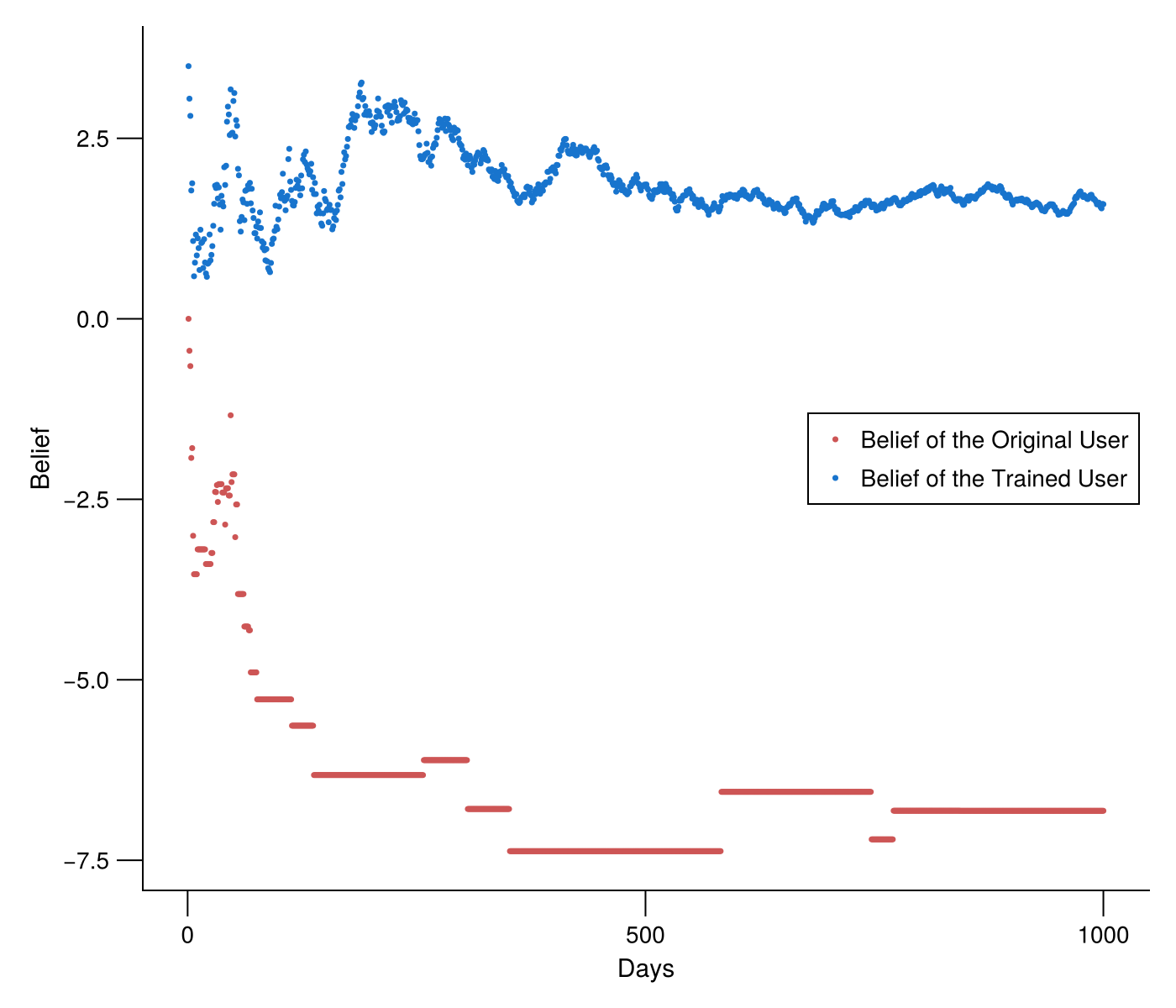}
    \caption{Belief Change Over Time}
    \label{fig:scatter}
  \end{minipage}

\end{figure}

\begin{figure}[ht]
  \centering
  \begin{minipage}{0.7\textwidth}
    \includegraphics[width=\textwidth]{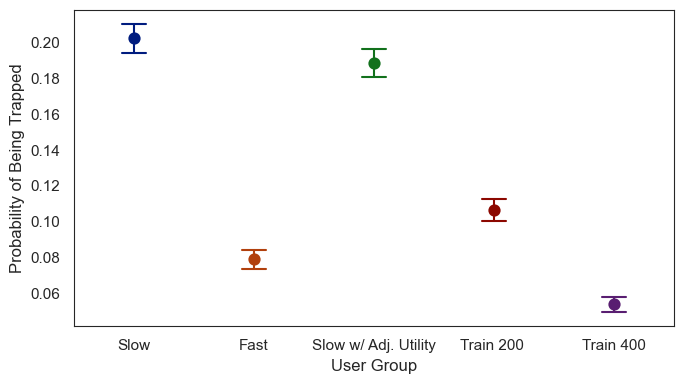}
    \caption{Simulated Probability of Being Trapped with Training}
    \label{fig:error_bar}
  \end{minipage}
\end{figure}

\begin{figure}[ht]
  \centering
  \begin{minipage}{0.7\textwidth}
    \includegraphics[width=\textwidth]{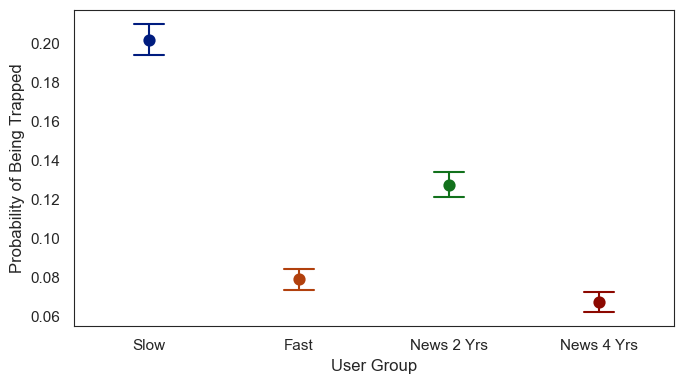}
    \caption{Simulated Probability of Being Trapped with Reading News}
    \label{fig:error_bar_news}
  \end{minipage}
\end{figure}

\begin{figure}[ht]
  \centering
  \begin{minipage}{1.0\textwidth}
    \includegraphics[width=\textwidth]{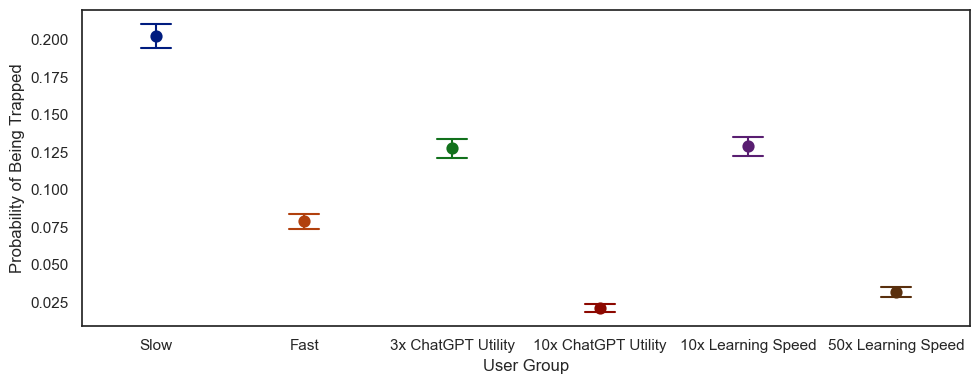}
    \caption{Simulated Probability of Being Trapped with Upgraded Models}
    \label{fig:error_bar_upgrade}
  \end{minipage}
\end{figure}

\end{APPENDICES}
\end{document}